\documentclass[12pt]{article}
\pdfoutput=1
\usepackage{amssymb,cite,graphicx}
\usepackage{slashed}
\usepackage{amsmath,bm,bbm}
\usepackage{amsfonts}
\usepackage[titletoc,title]{appendix}
\usepackage[small]{caption}
\usepackage[margin=1in]{geometry}
\usepackage[multiple]{footmisc}
\usepackage{mathtools}
\usepackage{slashed}
\usepackage[nottoc]{tocbibind}
\usepackage{xcolor}

\definecolor{mediumblue}{rgb}{0,0,0.8}
\usepackage{hyperref}
\hypersetup{
  linktocpage=true,
  colorlinks=true,
  citecolor=mediumblue,
  filecolor=mediumblue,
  linkcolor=mediumblue,
  urlcolor=mediumblue,
}

\graphicspath{{./}{./figures/}}
\numberwithin{equation}{section}
\allowdisplaybreaks%

\newcommand{\be}{\begin{equation}}
\newcommand{\ee}{\end{equation}}
\newcommand{\bea}{\begin{eqnarray}}
\newcommand{\eea}{\end{eqnarray}}
\begin{document}

\begin{titlepage}
  \begin{flushright}
TUM-HEP-1296/20
  \end{flushright}
  \medskip

  \begin{center}
    {\Large\bf\boldmath
    Exothermic dark mesons in light of electron recoil excess at XENON1T \vspace{0.3cm} \\
    }
\vspace{1.5cm}

    {\bf Soo-Min Choi$^{\dagger,1}$, Hyun Min Lee$^{\ddagger,2}$ and Bin Zhu$^{*,3,2}$ }

         {\it $^1$Physik Department T31, James-Franck-Stra\ss e 1, Technische Universit\"at M\"unchen, D-85748 Garching, Germany
         }\\[0.2cm]
    {\it $^2$Department of Physics, Chung-Ang University, Seoul
      06974, Korea}\\[0.2cm]
{\it $^3$School of Physics, Yantai University, Yantai 264005, China}
  \end{center}

  \bigskip

  \begin{abstract}
    \noindent
    We consider a novel mechanism to realize exothermic dark matter with dark mesons in the limit of approximate flavor symmetry in a dark QCD. We introduce a local dark $U(1)'$ symmetry  to communicate between dark mesons and the Standard Model via $Z'$ portal by partially gauging the dark flavor symmetry with flavor-dependent charges for cancelling chiral anomalies in the dark sector.  After the dark local $U(1)'$ is broken spontaneously by the VEV of a dark Higgs, there appear small mass splittings between dark quarks, consequently, leading to small split masses for dark mesons, required to explain the electron recoil excess in XENON1T by the inelastic scattering between dark mesons and electron. We propose a concrete benchmark model for split dark mesons based on $SU(3)_L\times SU(3)_R/ SU(3)_V$ flavor symmetry  and $SU(N_c)$ color group and show that there exists a parameter space making a better fit to the XENON1T data with two correlated peaks from exothermic processes and satisfying the correct relic density, current experimental and theoretical constraints.

  \end{abstract}

\vspace{4.5cm}
  \begin{flushleft}
     $^\dagger$Email: soo-min.choi@tum.de \\
    $^\ddagger$Email: hminlee@cau.ac.kr  \\
       $^*$Email: zhubin@mail.nankai.edu.cn
  \end{flushleft}
\end{titlepage}
 
\section{Introduction}

%Motive
Dark matter has provided an important playground for model building for new physics, due to the lack of elementary particles or composite analogues for dark matter in the Standard Model (SM). Weakly Interacting Massive Particles (WIMPs) have been best studied due to the testability in terrestrial experiments through their sizable interactions to the SM. However, the null signals for WIMP dark matter in various direct and indirect detection experiments lead us to ponder about alternative possibilities including light dark matter candidates below GeV scale with feeble interactions.

%Work
Recently there have been intriguing anomalies below $7\,{\rm keV}$ in the electron recoil energy reported by XENON1T experiment, at about $3\sigma$ deviation from the known background model \cite{xenon}.  There is still a need of understanding the background candidates for the anomalies such as tritium and accumulating more data with a longer period of time before making a definite conclusion on the XENON1T excess. Nonetheless,  it is worthwhile to pursue a consistent model for dark matter to explain the XENON1T excess.  Exothermic dark matter has drawn new attention in this regard, due to  the fact that a heavier component dark matter scatters off  electron  down to a lighter component dark matter \cite{exothermic}, producing the recoil energy of electron peaked at the mass splitting between two components, even for the standard Maxwellian distribution of dark matter velocity. Pseudo-Dirac fermion dark matter \cite{exodm} or complex scalar dark matter \cite{harigaya} with a dark $U(1)'$ symmetry have been proposed to explain the XENON1T excess. In these scenarios for exothermic dark matter, it is common to take dark matter and mediator particles to be light below GeV scale \cite{essig,exodmothers}, belonging to the category that has been of growing interest in recent years as alternatives to WIMP dark matter.

%Result
In this article, we propose a novel mechanism for exothermic dark matter based on the dark flavor symmetry $SU(N_f)_L\times SU(N_f)_R/SU(N_f)_V$  for dark quarks in $SU(N_c)$ dark QCD. In this scenario, dark mesons are bound states formed from dark quarks, and they are regarded naturally as candidates for light dark matter, thanks to the pseudo-Goldstone nature after the dark flavor symmetry is spontaneously broken by dark QCD condensation \cite{review}. We partially gauge the dark flavor symmetry by a dark  local $U(1)'$ to communicate between dark mesons and the SM via $Z'$ portal and take the $U(1)'$ charges of dark quarks to be vector-like but flavor-dependent  under the $U(1)'$ \cite{simpmeson2}.  Even if the dark flavor symmetry is broken explicitly by the flavor-dependent $U(1)'$,  the longevity of dark mesons is ensured due to the appropriate assignment of dark charges for no chiral anomalies in the dark sector\cite{simpmeson2}, as well as the approximate dark flavor symmetry with a small $U(1)'$ gauge coupling and a $Z'$ mass larger than dark QCD condensation scale.

In this work, we investigate a new origin of small mass splitting for exothermic dark mesons from small mixing Yukawa couplings between dark quarks and dark Higgs. After the local $U(1)'$ is broken by the VEV of a dark Higgs, small mixing masses for dark quarks are induced, giving rise to split masses for dark mesons and at the same time the meson-changing interactions for $Z'$, due to the fact that the mass matrix for dark quarks and the dark charge operator are not simultaneously diagonalized. We make a concrete discussion on the mass splitting and meson-changing interactions for $N_f=3$ case and search for a consistent parameter space for explaining the XENON1T excess with a better fit by two  correlated monochromatic peaks from exothermic dark mesons and satisfying various experimental constraints as well as the requirement from the correct relic density and the radiative stability of  dark mesons.

%Organize
The paper is organized as follows. 
We begin with a description of the model setup for dark mesons and introduce necessary interactions for the later discussion. Then, we show how the mass splitting between dark mesons is generated for $N_f=2$ and $N_f=3$ cases and also discuss the effects from $Z'$ gauge interactions on that.  Next, we collect the model-independent ingredients for split dark mesons  with $N_f>2$ in direct detection, relic density calculations, kinetic decoupling, late chemical decoupling and lifetime of dark mesons, in order.
Focusing on $N_f=3$ case, we continue to show the consistent parameter space for exothermic dark mesons in light of the XENON1T excess.
Finally, conclusions are drawn.
There is one appendix showing the details for Wess-Zumino-Witten interactions for $Z'$ and dark mesons for $N_f=3$.

\section{The setup}

We consider dark mesons as light dark matter living on $SU(N_f)_L\times SU(N_f)_R/SU(N_f)_V$ flavor symmetry and $SU(N_c)$ color group in the dark sector. A nonzero Wess-Zumino-Witten (WZW) term \cite{wz,witten} exists only for a nontrivial homotopy group, $\pi_5(G/H)=Z$, i.e. $N_f\geq 3$ for $G=SU(N_f)_L\times SU(N_f)_R$ and $H=SU(N_f)_V$. Then, general dark mesons can be described by the chiral perturbation theory with the WZW term in the dark sector \cite{review,simpmeson1,simpmeson2,simpmeson3,gori,split}.

We make a partial gauging of the flavor symmetry with a dark local $U(1)'$ and impose dark quarks to be vector-like under the $U(1)'$.   In order to protect neutral dark mesons from being decaying, we need to cancel the Axial-Vector-Vector current anomalies, simply chiral anomalies, for the dark chiral symmetry, by choosing the charge operator $Q'$ for dark quarks to satisfy ${\rm Tr}(Q^{\prime 2}t^a)=0$ for $t^a\in SU(N_f)_L\times SU(N_f)_R/SU(N_f)_V$ being broken generators of the flavor symmetry. To this, we can choose a simple but nontrivial choice, $Q^{\prime 2}=1$, so the charge operator takes $+1$ or $-1$ with ${\rm Tr}(Q')\neq 0$ \cite{simpmeson2}. In this case, some of dark mesons become charged under the $U(1)'$. We assume that the dark $U(1)'$ is broken spontaneously due to the VEV of a dark Higgs $\phi$, so the corresponding dark gauge boson $Z'$ gets massive.

The Lagrangian for dark mesons $\pi$, dark gauge boson $Z'$, and dark Higgs $\phi$ are given by
\bea
{\cal L}&=& -\frac{1}{4} F'_{\mu\nu} F^{\prime\mu\nu} -\frac{1}{2} \sin\xi\, F'_{\mu\nu} B^{\mu\nu}+|D_\mu\phi|^2-V(\phi) \nonumber \\
&&+ \frac{f^2_\pi}{4} {\rm Tr} \Big[D_\mu\Sigma (D^\mu \Sigma )^\dagger\Big ] +\frac{f^2_\pi}{2}{\rm Tr} \Big[\mu (M\Sigma+ \Sigma^\dagger M) \Big] +\frac{ f^4_\pi}{4} c\, g^2_{Z'}  {\rm Tr}\Big[Q'\Sigma Q' \Sigma^\dagger \Big]  \nonumber \\
&&+{\cal L}_{WZW} +{\cal L}_{gWZW} \label{Lagfull}
\eea
where $F'_{\mu\nu}=\partial_\mu Z'_\nu-\partial_\nu Z'_\mu$ is the field strength tensor, $\Sigma={\rm exp}(i2\pi/f_\pi)$ with $\pi=\pi^a t^a$ contains the dark mesons, 
and the covariant derivatives are $D_\mu\phi=(\partial_\mu+ i q_\phi g_{Z'} Z'_\mu)\phi$ with $U(1)'$ charge being $q_\phi=+2$ and $D_\mu \Sigma=\partial_\mu\Sigma +i g_{Z'} [Q',\Sigma] Z'_\mu$. Here, $\xi$ is the gauge kinetic mixing, $M$ is the dark quark mass matrix,  $\mu$ parametrizes the dark QCD condensation scale, and the coefficient of the $Z'$ corrections to dark mesons is parametrized as $c\sim \frac{1}{16\pi^2} \,\frac{\mu^2}{m^2_{Z'}}$ \cite{split}.
We also note that ${\cal L}_{WZW}$ is the Wess-Zumino-Witten term, which contains in the leading order,
\bea
{\cal L}_{WZW}=\frac{2 N_c}{15\pi^2 f^5_\pi} \,\epsilon^{\mu\nu\rho\sigma} {\rm Tr}[ \pi \partial_\mu\pi \partial_\nu \pi \partial_\rho\pi \partial_\sigma \pi], \label{WZW}
\eea
and ${\cal L}_{gWZW}$ includes the additional $Z'$ invariant meson interactions \cite{witten}, which are given  in the leading order by
\bea
{\cal L}_{gWZW}=\frac{i N_c g_{Z'}}{3\pi^2 f^3_\pi}  \epsilon^{\mu\nu\rho\sigma}\, Z'_\mu\,  {\rm Tr} [Q' \partial_\nu\pi \partial_\rho\pi \partial_\sigma \pi].  \label{gWZW}
\eea
Here, we find that there is no extra $Z'-Z'-\pi-\pi$ coupling  coming from the gauged WZW term in our case due to $Q^{\prime 2}=1$.  The WZW term is important for determining the relic density with $3\to 2$ annihilation processes for SIMP dark mesons with $N_f\geq 3$ in the strongly coupled regime \cite{simpmeson1,simpmeson2,simpmeson3,split}.

From the chiral Lagrangian in the dark sector, we obtain the quartic self-interactions for dark mesons as follows,
\bea
{\cal L}_{\pi, 4} = \frac{2}{3f^2_\pi}\, {\rm Tr} [(\partial_\mu\pi) \pi (\partial^\mu\pi)\pi-\pi^2 (\partial_\mu\pi)(\partial^\mu\pi)]. \label{meson4}
\eea
The dark meson mass terms in the full Lagrangian in eq.~(\ref{Lagfull}) give rise to extra quartic self-interactions for dark mesons.

The dark Higgs potential $V(\phi)$ takes the form, $V(\phi)=m^2_\phi |\phi|^2 +\lambda_\phi |\phi|^4 $, and the dark Higgs can mix with the SM Higgs by a quartic coupling, $-\lambda_{\phi H} |\phi|^2 |H|^2$.
After the dark Higgs is expanded around a nonzero VEV by $\langle\phi\rangle=(v_\phi+h')/\sqrt{2}$, $Z'$ and  the dark Higgs masses are given by $m_{Z'}=q_\phi g_{Z'} v_\phi$ and $m_{h'}=\sqrt{2\lambda_\phi}  v_\phi$, in the limit of small mixings with the visible sector. Moreover, the dark Higgs interactions to $Z'$ are also given by
\bea
{\cal L}_{h',{\rm int}} = \frac{m^2_{Z'}}{v_\phi}\, h' Z'_{\mu}   Z^{\prime\mu} +\frac{1}{2} q^2_\phi g^2_{Z'} h^{\prime 2}  Z'_{\mu}   Z^{\prime\mu}.
\eea
The dark Higgs also has small mixing Yukawa couplings to dark quarks as will be discussed in the next section, but the resulting dark Higgs interactions to dark mesons are suppressed for small mass splittings for dark mesons.
Moreover, the dark Higgs can have a small mixing with the SM Higgs through the Higgs portal coupling $\lambda_{\phi H}$, but we assume it to be small enough to satisfy the phenomenological bounds such as Higgs invisible decay but it can be nonzero for kinetic equilibrium with the SM during the freeze-out.

There is a communication between dark matter and the SM, due to the  gauge kinetic term between the dark photon $Z'$ and the SM hypercharge \cite{simpmeson2}. 
For a small gauge kinetic mixing in eq.~(\ref{Lagfull}), that is, $\xi\ll 1$, the $Z'$ interactions to the SM are approximated \cite{z3dm,exodm} by
\bea
{\cal L}_{Z',{\rm SM}}=- e\varepsilon Z'_\mu \bigg( J^\mu_{\rm EM}+\frac{m^2_{Z'}}{2c^2_W m^2_Z}\, J^\mu_Z  \bigg)+g_{Z'} Z'_\mu J^\mu_{Z'}
\eea
where $\varepsilon\equiv c_W \xi$ with $c_W=\cos\theta_W$, and  $J^\mu_{\rm EM}, J^\mu_Z$ are electromagnetic and neutral currents in the SM, for instance, $J^\mu_{\rm EM}={\bar e} \gamma^\mu e$ for electron and $J^\mu_{Z}={\bar \nu} \gamma^\mu P_L \nu$ for neutrinos, and $J^\mu_{Z'}$ is the dark $U(1)'$ current.
Then, dark mesons can scatter off the electron through $Z'$-portal for direct detection of dark matter, and dark mesons can pair annihilate into $e^+e^-$ for determining the relic density.

\section{Mass splittings and flavor violation for dark mesons}

If the charge operator for $U(1)'$ is not universal,   in general, the $Z'$ gauge interactions to dark quarks do not remain diagonal in the basis of mass eigenstates after the mass matrix for dark quarks is diagonalized.  Suppose that the mass matrix for dark quarks is diagonalized to
\bea
M_{\rm diag} = V_L M V^\dagger_R = {\rm diag} (m'_1, m'_2, \cdots,m'_{N_f})
\eea
with $V_L$ and $V_R$ being rotation matrices for left-handed and right-handed dark quarks. 
As a result, the mass terms for dark mesons become
\bea
{\cal L}_m =-\frac{f^2_\pi}{2}{\rm Tr} \Big[\mu (M_{\rm diag}{\widetilde\Sigma}+ {\widetilde\Sigma}^\dagger M_{\rm diag}) \Big]
-\frac{ f^4_\pi}{4} c\, g^2_{Z'} {\rm Tr}\Big[(V_R Q' V^\dagger_R) {\widetilde\Sigma} (V_L Q' V^\dagger_L)  {\widetilde\Sigma}^\dagger \Big]  \label{mesonmass}
\eea
with the mesons being redefined as 
\bea
{\widetilde\Sigma}=V_R \Sigma V^\dagger_L\equiv {\rm exp(i2{\widetilde\pi}/f_\pi)}.
\eea
Then, the meson mass terms can be identified from the expansion of the first term in eq.~(\ref{mesonmass}) and they receive corrections from the $Z'$ interactions in the second term of eq.~(\ref{mesonmass}).

On the other hand, the covariant derivative for redefined dark mesons becomes, in the basis of the diagonalized mass matrix,
\bea
D_\mu {\widetilde\Sigma}=\partial_\mu{\widetilde\Sigma} +i g_{Z'} (V_R Q'  V^\dagger_R {\widetilde\Sigma}-{\widetilde\Sigma} V_L Q' V^\dagger_L) Z'_\mu. \label{Zpint2}
\eea
Therefore, for the flavor-dependent $Q'$, the new charge operators, $V_R Q' V^\dagger_R$ or $V_L Q' V^\dagger_L$, appearing in the covariant derivatives for dark mesons, do not have to be flavor diagonal\footnote{Split masses for dark mesons were also discussed in Ref.~\cite{split}, but the dark charge operator for $Z'$ and the dark quark mass matrix are taken to be diagonalized simultaneously, unlike our case. }, leading to flavor-changing meson interactions with $Z'$.
However, the quartic self-interactions in eq.~(\ref{meson4}) and WZW terms  in eq.~(\ref{WZW}) for the redefined dark mesons take the same forms as for the original dark mesons.

Suppose that the mass matrix for dark quarks have the degenerate diagonal entries $m_1$, due to the $SU(N_f)$ flavor symmetry.
Introducing the dark charge operator of the form \cite{simpmeson2},
\bea
Q'={\rm diag}(+1,-1,-1,\cdots,-1),
\eea 
that is, $q'_1$ quark carries the opposite $U(1)'$ charge from those of $q'_j$ with $j=2,3,\cdots,N_f$,
we can write down the following the mixing Yukawa couplings between dark quarks and the dark Higgs $\phi$ carrying $+2$ charge,
\bea
{\cal L}_{\rm mix} = - \sum_{i\neq j} y_{ij}\, \phi \,{\bar q}'_i q'_j - {\rm h.c.}
\eea
Then, after the $U(1)'$ symmetry is broken spontaneously due to a nonzero VEV of $\phi$,  the mixing mass terms are generated, so  the $SU(N_f)$ flavor symmetry is broken explicitly.
For small Yukawa couplings, we can generate small mass mixing parameters, $y_{ij}\langle\phi\rangle\ll m_1$, so the mass splittings between dark quarks appear small.

\subsection{$N_f=2$ case}

For $N_f=2$, the dark mesons take the following form,
\bea
\pi =\frac{1}{\sqrt{2}} \left(\begin{array}{cc} \frac{1}{\sqrt{2}} \pi^0   & \pi^+ \vspace{0.2cm}  \\  
\pi^-& -\frac{1}{\sqrt{2}} \pi^0  \end{array} \right).
\eea
Choosing the dark charge operator as
\bea
Q'=\left(\begin{array}{cc} 1& 0   \\ 0 & -1   \end{array} \right), \label{charge2}
\eea
and assuming that the mass matrix for dark quarks is diagonal,
we obtain the $Z'$ gauge interations as
\bea
{\cal L}_{Z',2\pi}=2i g_{Z'} Z'_\mu (\pi^+\partial_\mu \pi^- -\pi^-\partial_\mu\pi^+) +4 g^2_{Z'} Z'_\mu Z^{\prime\mu} \pi^+\pi^-.\label{zp-n2}
\eea
In this case, no WZW term is allowed, so neither is gauged counterpart for $Z'$.

In the limit of vanishing $Z'$ corrections, the dark meson masses for $N_f=2$ are given by
\bea
m^2_{{\widetilde\pi}^\pm} =m^2_{{\tilde\pi}^0}= \mu (m'_1+m'_2).
\eea
We take the dark quark mass matrix to be deviated by identity due to nonzero off-diagonal components, as follows,
\bea
M= \left( \begin{array}{cc}  m_1 & \epsilon \\ \epsilon & m_1  \end{array}\right)
\eea
with $\epsilon=y_{12} \langle\phi\rangle$.
Then, the above mass matrix is diagonalized by
\bea
V_R=V_L =  \left( \begin{array}{cc}  \frac{1}{\sqrt{2}} &  - \frac{1}{\sqrt{2}}\vspace{0.2cm} \\   \frac{1}{\sqrt{2}}  &  \frac{1}{\sqrt{2}} \end{array}\right),
\eea
and the mass eigenvalues are given by
\bea
m'_1 &=& m_1-\epsilon, \\
m'_2 &=& m_1+\epsilon.
\eea
So, in this case, $m^2_{{\widetilde\pi}^\pm} =m^2_{{\widetilde\pi}^0}=2\mu m_1$, so there is no mass splitting from the mass mixing of dark quarks.
However, the $Z'$ interactions make the meson mass splitting, as follows,
\bea
m^2_{{\widetilde\pi}^1} &=&m^2_{\widetilde{\pi}}, \\  
m^2_{{\widetilde\pi}^0} &=& m^2_{{\widetilde\pi}^2} =m^2_{\widetilde{\pi}}-2\delta  \label{p2} 
\eea
with $m^2_{\tilde\pi}=2\mu m_1$, $\delta=c g^2_{Z'} f^2_\pi$, and ${\widetilde\pi}^\pm =\frac{1}{\sqrt{2}}({\widetilde\pi}^1\mp i{\widetilde\pi}^2)$.
Then, for $\delta \ll m^2_{\widetilde{\pi}}$, the mass difference is given by $m_{{\widetilde\pi}^{1}}-m_{{\widetilde\pi}^{0,2}}\simeq \delta/m_{\tilde\pi}$.  For $N_f=2$, it is crucial to include the $Z'$ corrections to the meson mass splitting.

As compared to eq.~(\ref{zp-n2}), the $Z'$ interactions are maintained in the basis of mass eigenstates,
\bea
{\cal L}_{Z',{\rm int}} =2 g_{Z'} Z'_\mu  ({\tilde\pi}^2\partial^\mu {\tilde\pi}^0-{\tilde\pi}^0\partial^\mu {\tilde\pi}^2)  + 2g^2_{Z'} Z'_\mu Z^{\prime\mu} [({\widetilde\pi}^0)^2+({\widetilde\pi}^2)^2 ].
\eea
In this case, there is no exothermic process through $Z'$ interactions, because ${\tilde\pi}^0, {\tilde\pi}^2$ are still mass degenerate.
Therefore, we need to go beyond $N_f=2$ to realize a minimal scenario for exothermic dark mesons through $Z'$ interactions, so we focus on the $N_f=3$ case in the following discussion. Nonetheless, it is still interesting to consider a minimal dark matter scenario where  dark mesons with split masses are self-interacting and have $Z'$ portal interactions.

\subsection{$N_f=3$ case}

For $N_f=3$, there are additional dark mesons given in the following form,
\bea
\pi =\frac{1}{\sqrt{2}} \left(\begin{array}{ccc} \frac{1}{\sqrt{2}} \pi^0+\frac{1}{\sqrt{6}}\eta^0   & \pi^+ & K^+ \vspace{0.2cm} \\  
\pi^-& -\frac{1}{\sqrt{2}} \pi^0+\frac{1}{\sqrt{6}}\eta^0   & K^0 \vspace{0.2cm}\\  K^- & \overline{K^0} & -\frac{2}{\sqrt{6}}\eta^0 \end{array} \right).
\eea
Taking the charge operator $Q'$ for dark quarks under the $U(1)'$ \cite{simpmeson2} to be
\bea
Q'=\left(\begin{array}{ccc} 1& 0 & 0 \\ 0 & -1 & 0 \\ 0 & 0 & -1 \end{array} \right), \label{charge}
\eea 
and 
assuming that the mass matrix for dark quarks is diagonal,
we determine the $Z'$ gauge interactions \cite{simpmeson2} by
\bea
{\cal L}_{Z',2\pi} &=& 2i g_{Z'} Z'_\mu \Big(K^+\partial^\mu K^- -K^- \partial^\mu K^+ + \pi^+ \partial^\mu \pi^- -  \pi^- \partial^\mu \pi^+ \Big) \nonumber \\
&&+ 4 g^2_{Z'}Z'_\mu Z^{\prime \mu} (K^+ K^- + \pi^+ \pi^-). \label{zpdm}
\eea
We also note that from eq.~(\ref{gWZW}), the gauged WZW terms contain the $Z'$ couplings to three dark mesons for  the dark charge operator in eq.~(\ref{charge}). These cubic dark meson interactions to $Z'$ are important for determining the relic abundance from the semi-annihilation of dark mesons, $\pi^i\pi^j\rightarrow \pi^k Z' $.

In the limit of vanishing $Z'$ corrections, the dark meson masses for $N_f=3$ are given by
\bea
m^2_{{\widetilde\pi}^\pm} &=& \mu (m'_1+m'_2), \\
m^2_{{\widetilde K}^\pm} &=& \mu (m_1'+m'_3), \\
m^2_{{\widetilde K}^0} &=& \mu (m'_2 + m'_3),
\eea
and ${\widetilde\pi}^0,{\widetilde\eta}^0$ mix by the following mixing mass matrix,
\bea
M^2_0 = \mu \left(\begin{array}{cc} m'_1+m'_2 & \frac{1}{\sqrt{3}}(m'_1-m'_2) \\   \frac{1}{\sqrt{3}}(m'_1-m'_2) &  \frac{1}{3}(m'_1 + m'_2 + 4m'_3) \end{array} \right).
\eea
Then, in the limit of degenerate masses for dark quarks, i.e. $m'_1=m'_2=m'_3$, all the dark mesons have the same masses as $m^2_{\widetilde{\pi}}=2\mu m'_1$. 
But, for non-degenerate masses for dark quarks, dark meson masses are not degenerate any longer. 

Including the mixing mass terms for dark quarks, we take the mass matrix for dark quarks to be deviated from identity, as follows,
\bea
M= \left( \begin{array}{ccc}  m_1 & \epsilon  & \epsilon' \\ \epsilon & m_1 &  0  \\ \epsilon' &  0 & m_1  \end{array}\right)
\eea
with $\epsilon=y_{12} \langle\phi\rangle$ and $\epsilon'=y_{13} \langle\phi\rangle$.
We note that the $(2,3), (3,2)$ entries in the dark quark mass matrix are taken to zero, because we assumed that the $SU(3)$ flavor symmetry is restored for the unbroken $U(1)'$.

Due to the violation of flavor symmetry with $\epsilon\neq 0$ and $\epsilon'\neq 0$, the rotation matrices are nontrivial and they are given by
\bea
V_R=V_L =  \left( \begin{array}{ccc}  \frac{1}{\sqrt{2}} &  - \frac{1}{\sqrt{2}}\cos\theta &- \frac{1}{\sqrt{2}}\sin\theta \vspace{0.2cm} \\   \frac{1}{\sqrt{2}}  &  \frac{1}{\sqrt{2}}\cos\theta  &  \frac{1}{\sqrt{2}}\sin\theta \vspace{0.2cm} \\ 0 &  -\sin\theta  &  \cos\theta \end{array}\right),
\eea
with
\bea
\sin\theta\equiv \frac{\epsilon'}{\sqrt{\epsilon^2+\epsilon^{\prime 2}}},
\eea
and the mass eigenvalues are given by
\bea
m'_1 &=& m_1-\sqrt{\epsilon^2+\epsilon^{\prime 2}}, \\
m'_2 &=& m_1+\sqrt{\epsilon^2+\epsilon^{\prime 2}}, \\
m'_3&=& m_1.
\eea
In this case, even without including $Z'$ interactions, we get the split dark meson masses,
\bea
m^2_{{\widetilde\pi}^\pm} &=&m^2_{\widetilde{\pi}} , \label{m1} \\
m^2_{{\widetilde\pi}^0} &=& m^2_{\widetilde{\pi}} \Big(1-\frac{2\Delta m }{\sqrt{3} m_{\widetilde{\pi}}} \Big),  \label{m2} \\
m^2_{{\widetilde K}^\pm} &=&m^2_{\widetilde{\pi}} \Big(1-\frac{\Delta m }{m_{\widetilde{\pi}}} \Big),  \label{m3}\\
m^2_{{\widetilde K}^0} &=&m^2_{\widetilde{\pi}} \Big(1+\frac{\Delta m }{m_{\widetilde{\pi}}} \Big),\label{m4} \\
m^2_{{\widetilde \eta}^0} &=&m^2_{\widetilde{\pi}} \Big(1 +\frac{2\Delta m }{\sqrt{3} m_{\widetilde{\pi}}} \Big).\label{m5}
\eea
where $m^2_{\widetilde{\pi}}\equiv 2\mu m_1$ and $\Delta m \equiv \mu \sqrt{\epsilon^2+\epsilon^{\prime 2}}/m_{\widetilde{\pi}}$,
Therefore, we get the mass hierarchy for dark mesons as $m_{{\widetilde\eta}^0}>m_{{\widetilde K}^0}>m_{{\widetilde\pi}^\pm}>m_{{\widetilde K}^\pm}>m_{{\widetilde\pi}^0}$.
The mass splittings for the dark mesons participating in the $Z'$ interactions are given by
\bea
m^2_{{\widetilde K}^0} -m^2_{{\widetilde K}^\pm} =\sqrt{3}\Big(m^2_{{\widetilde\pi}^\pm} -m^2_{{\widetilde\pi}^0}\Big) =2\Delta m.  \label{massrelation}
\eea
The result is different from the Dashen's mass relation \cite{dashen}, because the meson masses get split due to the mixing between three dark quarks, instead of the gauge corrections.
As a result, for $\Delta m \ll m_{\widetilde{\pi}}$ with small $\epsilon, \epsilon'$, the mass differences are given by $m_{{\widetilde K}^0}-m_{{\widetilde K}^\pm}\simeq\Delta m $  and $m_{{\widetilde\pi}^\pm}-m_{{\widetilde\pi}^0}\simeq \frac{1}{\sqrt{3}}\,\Delta m $. 
We remark that for $N_f=3$, there is no need of $Z'$ corrections for dark meson mass splitting, unlike the case with $N_f=2$.

As compared to eq.~(\ref{zpdm}),  the $Z'$ gauge interactions to redefined dark mesons with eq.~(\ref{Zpint2}) are given by
\bea
{\cal L}_{Z',{\rm int}} &=&i g_{Z'} Z'_\mu \bigg[ ({\widetilde K}^0+{\widetilde K}^+)\partial^\mu ( \overline{{\widetilde K}^0}+{\widetilde K}^-) -(\overline{{\widetilde K}^0}+ {\widetilde K}^-)\partial^\mu ({\widetilde K}^0+ {\widetilde K}^+)  \nonumber \\
&&\quad -\sqrt{2}  ({\widetilde \pi}^- - {\widetilde\pi}^+) \partial^\mu {\widetilde\pi}^0 +\sqrt{2}  {\widetilde\pi}^0 \partial^\mu ({\widetilde\pi}^- - {\widetilde\pi}^+) \bigg] \nonumber \\
&&+ g^2_{Z'}Z'_\mu Z^{\prime \mu} \bigg[ 2({\widetilde K}^0 + {\widetilde K}^+)(\overline{{\widetilde K}^0} + {\widetilde K}^-)+2({\widetilde \pi}^0)^2-({\widetilde \pi}^--{\widetilde \pi}^+)^2 \bigg]. \label{Zpint} 
\eea
As a consequence, there are not only flavor-conserving interactions for dark mesons but also flavor-changing interactions for dark mesons with split masses, such as ${\widetilde K}^0\rightarrow {\widetilde K}^+$, ${\widetilde\pi}^\pm \rightarrow {\widetilde\pi}^0$,  thus realizing the exothermic dark matter for an appropriate mass splitting between dark mesons.
Thus, the exothermic processes for non-degenerate dark mesons can be responsible for explaining the electron recoil excess in XENON1T experiment, while the elastic scattering processes for degenerate dark mesons can be tested in other light dark matter experiments.

We remark that the dark meson masses also receive radiative corrections due to $Z'$ interactions as in the case with $N_f=2$.  Then, masses for kaon-like and pion-like mesons get split further, but ${\tilde\eta}^0$ meson keeps the same mass as in eq.~(\ref{m5}).

First, for kaon-like dark mesons, the mass matrix in the general rotated basis of $({\tilde K}^0, {\tilde K^+})$ is corrected due to $Z'$ interactions to
\bea
M^2_{\tilde K} = \left(\begin{array}{cc}  m^2_{\tilde\pi} \Big(1+\frac{\Delta m}{ m_{\tilde\pi} }\Big)-\delta & -\delta  \\  -\delta &m^2_{\tilde\pi} \Big(1-\frac{\Delta m}{ m_{\tilde\pi} }\Big)-\delta  \end{array} \right)
\eea
where $\delta= c\, g^2_{Z'} f^2_\pi$.  Then, the mass eigenvalues for the kaon-like mesons become
\bea
m^2_{K_{1,2}} = m^2_{\tilde\pi} -\delta \pm \sqrt{\delta^2 + m^2_{\tilde\pi}(\Delta m)^2}. \label{kaonmassf}
\eea
So, if $\delta\gtrsim m_{\tilde\pi} \Delta m$, we would get $m^2_{K_{1}}\simeq  m^2_{\tilde\pi} $ and  $m^2_{K_{2}}\simeq  m^2_{\tilde\pi}-2\delta $, for which the mass splitting is given dominantly by the $Z'$ corrections as $m^2_{K_{1}}-m^2_{K_{2}}\simeq 2\delta$.
However, we can maintain $m_{{\widetilde K}^0}-m_{{\widetilde K}^\pm}\simeq \Delta m$ for $\delta\lesssim m_{\tilde\pi} \Delta m$, that is, if the following condition is satisfied,
\bea
m_{Z'}\gtrsim 0.6\mu\,\Big(\frac{g_{Z'}}{0.01}\Big)\bigg(\frac{0.2}{m_{\tilde\pi}/f_\pi}\bigg)\bigg(\frac{m_{\tilde\pi}/100\,{\rm MeV}}{\Delta m/4\,{\rm keV}}\bigg)^{1/2}.
\eea 
Here, the dark QCD condensation scale $\mu$ is constrained by dark meson mass $m_{\tilde\pi}$ and dark quark mass $m_1$ to be $\mu=m^2_{\tilde\pi}/(2m_1)$.

Similarly, the $Z'$ gauge interactions also make the masses for pion-like dark mesons split, whose mass matrix is, in the basis of $({\widetilde\pi}^0, {\widetilde\pi}^1=\frac{1}{\sqrt{2}}({\widetilde \pi}^-+{\widetilde \pi}^+), {\widetilde\pi}^2=\frac{1}{\sqrt{2}i}({\widetilde \pi}^--{\widetilde \pi}^+))$, given by
\bea
m^2_{{\tilde\pi}^0} &=& m^2_{\tilde\pi} \Big(1-\frac{2\Delta m}{ \sqrt{3}m_{\tilde\pi}} \Big) -2\delta,  \label{pionmassf1} \\
 m^2_{{\tilde\pi}^1} &=& m^2_{\tilde\pi}, \label{pionmassf2} \\
m^2_{{\tilde\pi}^2} &=&   m^2_{\tilde\pi}  -2\delta.  \label{pionmassf3}
\eea
Thus, from eq.~(\ref{Zpint}), the exothermic process between ${\widetilde\pi}^0$ and ${\widetilde\pi}^2$ through $Z'$ is subject to the $Z'$ mass correction.
But, as far as the $Z'$ gauge corrections are bounded similarly as for kaon-like dark mesons, the inelastic scattering between pion-like mesons and electron through $Z'$ can be still responsible for the XENON1T electron excess.

Consequently, from the results in eqs.~(\ref{kaonmassf}) and (\ref{pionmassf1})-(\ref{pionmassf3}) that the simultaneous presence of the mixing between dark quarks and the $Z$ corrections lead to the modified Dashen's relation for dark meson masses,
\bea
m^2_{K_1} -m^2_{K_2} =\sqrt{(m^2_{{\tilde\pi}^1}- m^2_{{\tilde\pi}^2})^2 + 3(m^2_{{\tilde\pi}^2}- m^2_{{\tilde\pi}^0})^2} 
\eea
with $ m^2_{{\tilde\pi}^1}- m^2_{{\tilde\pi}^2} =2\delta$ and $m^2_{{\tilde\pi}^2}- m^2_{{\tilde\pi}^0}=\frac{2}{\sqrt{3}}\,m_{\tilde\pi} \Delta m $. In general, the kaon-like dark mesons have the largest mass splitting. For $\Delta m=0$, we recover the Dashen's relation, $m^2_{K_1} -m^2_{K_2}=m^2_{{\tilde\pi}^1}- m^2_{{\tilde\pi}^2}$. For $\delta=0$, we recover the previous result in eq.~(\ref{massrelation}).

For the later discussion, we focus on the case with $\delta\ll m_{\tilde\pi} \Delta m$, thus we make use of the mass formulas in eqs.~(\ref{m1})-(\ref{m5}) and the $Z'$ interactions in eq.~(\ref{Zpint}).

\section{General discussion on split dark mesons }

In this section, we provide a general discussion on the phenomenology of split dark mesons that are applicable in a more general framework.
Dark mesons with split masses can give rise to exothermic processes for explaining the XENON1T electron recoil excess with more than one peaks.  Boltzmann equations for determining the relic density and kinetic decoupling conditions are presented. The crucial issues on late decoupling and lifetime of heavier dark mesons are also discussed.

\subsection{Dark mesons and XENON1T electron recoil}

As discussed in the previous section, in the presence of flavor violation in the dark sector, dark mesons get split masses and their mixings give rise to $Z'$ gauge interactions changing between dark mesons. 
As far as $m_{\widetilde{\pi}}\gtrsim 10\,{\rm MeV}$, we can ignore the velocity of the bound electrons in Xenon atoms \cite{exodm}, so we assume that this is the case in our discussion.

Ignoring the $Z'$ gauge corrections to the mass splittings for dark mesons, we have 
$m_{{\widetilde \pi}_i}-m_{{\widetilde \pi}_j}\simeq\Delta m_{ij}>0 $ due to the meson mixings.
Then, it is possible to realize the exothermic scattering process,  ${\widetilde \pi}_i e\rightarrow {\widetilde \pi}_j e$ \cite{exodm}.
Taking $\Delta m_{ij} \ll m_e\ll m_{\widetilde{\pi}}$ and 
\bea
\kappa_{ij}\equiv \frac{2\Delta m_{ij}}{m_e v^2} \gg 1,
\eea
we use the approximate formulas for the electron recoil energy and the momentum transfer for  ${\widetilde \pi}_i e\rightarrow {\widetilde \pi}_j e$ \cite{exodm}, as follows,
\bea
E_R
&\simeq&  \Delta m_{ij}  \bigg(1-\frac{2}{\sqrt{\kappa_{ij}}}\cos\theta \bigg), \\
q^2 &\simeq& 2m_e \Delta m_{ij} \bigg(1-\frac{2}{\sqrt{\kappa_{ij}}}\, \cos\theta \bigg) \label{qapprox}
\eea
where $\theta$ is the scattering angle between dark meson and electron in the center of mass frame.

Then, we get the total event rate per Xenon detector for dark mesons \cite{exodm} as
\bea
R_D 
&\simeq &50 \bigg(\frac{M_T}{\rm tonne-yrs}\bigg) \bigg(\frac{{\bar\sigma}_e/m_{\tilde\pi}}{1.2\times 10^{-43}\,{\rm cm}^2/{\rm GeV}} \bigg)  \nonumber \\
&&\times \sum_i\bigg(\frac{ r_i \,\rho_{{\widetilde \pi}_i} \, K_{\rm int}(\Delta m_{ij}\big)}{2.6\cdot(0.4\,{\rm GeV\, cm^{-3}})} \bigg)  \bigg(\frac{\Delta m_{ij}}{2.5\,{\rm keV}}\bigg)^{1/2} \label{total}
\eea
where $m_{\widetilde{\pi}}= 2\mu m_1$ is the common dark meson mass in the limit of the unbroken flavor symmetry.
Here, for $m_e, m_{\widetilde{\pi}}, m_{Z'}\gg q\simeq \sqrt{2m_e \Delta m}$, the scattering cross section between the dark meson and electron is normalized to the elastic scattering cross section with $\Delta m_{ij}=0$, as follows,
\bea
{\bar\sigma}_e\simeq  \frac{\varepsilon^2 e^2 g^2_{Z'}\mu^2_{e\pi}}{\pi m^4_{Z'}},
\eea
with $\mu_{e\pi}=m_e m_{\widetilde{\pi}}/(m_e+m_{\widetilde{\pi}})$ being the reduced mass for dark meson-electron system, $r_i$ denotes the inelastic scattering cross section for ${\widetilde \pi}_i$ in units of ${\bar\sigma}_e$,
$M_T$ is the fiducial mass of the detector, given by $M_T\simeq 4.2\times 10^{27}(M_T/{\rm tonne}) m_T$ for Xenon, $K_{\rm int}(E_R)$ is the integrated atomic excitation factor  normalized to $E_R=2.5\,{\rm keV}$,  and $ \rho_{{\widetilde\pi}_i}$ are the local energy densities of dark mesons.

We note that the integrated atomic excitation factor introduced in eqs~(\ref{total}) or (\ref{totaln3}) is given by
\bea
K_{\rm int} (E_R,q) =\int^{q_+}_{q_-} a^2_0\, q\, dq\, K(E_R,q)
\eea
where $q_\pm$ are the maximum and minimum values of the momentum transfer, $a_0$ is the Bohr radius,
\bea
K(E_R,q) =\frac{\alpha^2 m_e (m_e+m_{\chi_1})^2}{4E_R m^2_{\chi_1}} \sum_{n,l} |f_{nl}^{\rm ion}(p_e,q)|^2 
\eea
with $p_e=\sqrt{2m_e E_R}$ being the outgoing momentum of electron and $\alpha$ being the fine structure constant, and  $f_{nl}^{\rm ion}(p_e,q)$ is the non-relativistic ionization form factor.
 
 \begin{figure}[tbp]
  \centering
  \includegraphics[width=.50\textwidth]{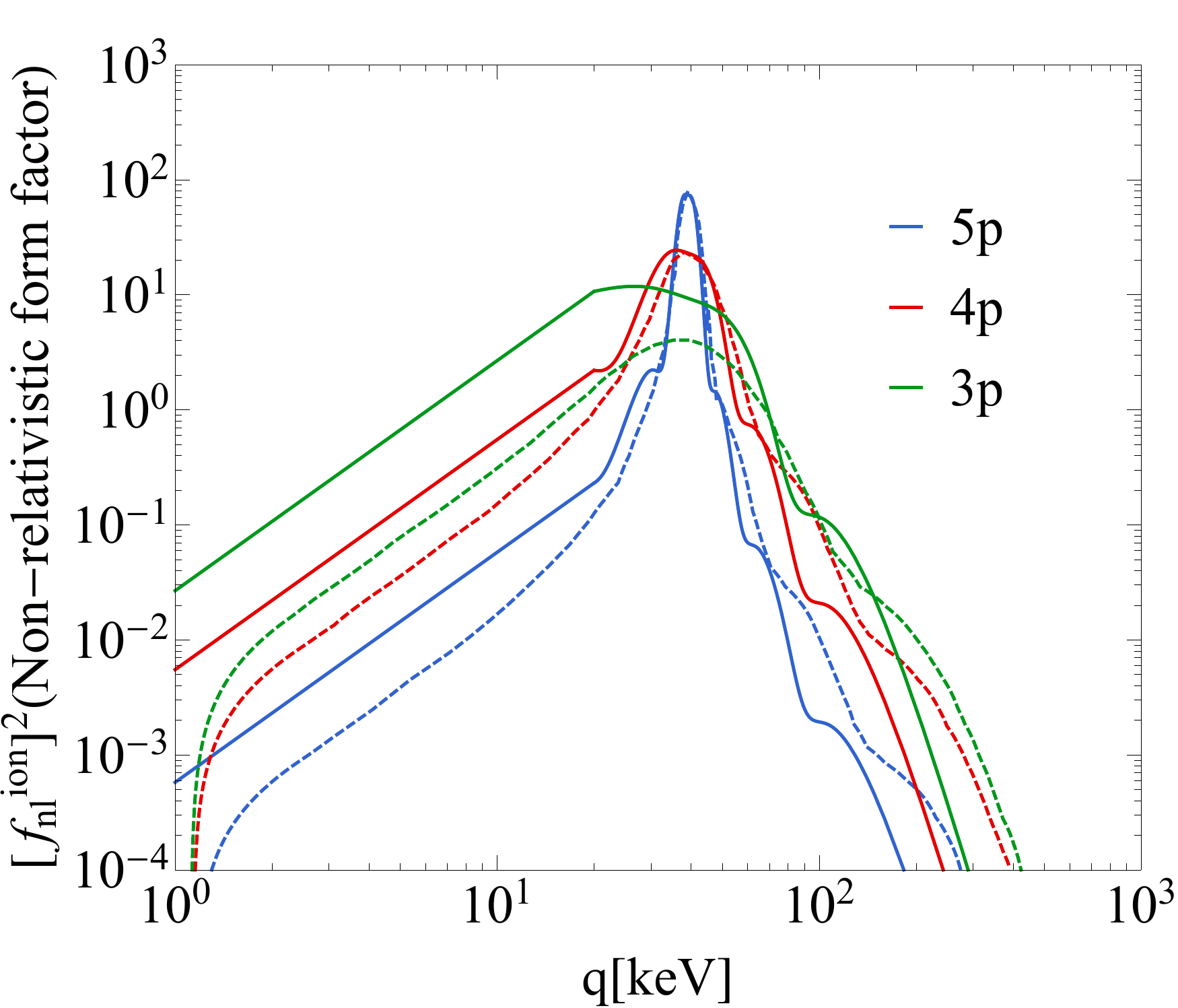}   
  \caption{Ionization  form factor as a function of momentum transfer $q$ in keV for $5p, 4p, 3p$ shells in blue, red and green lines. Our results are shown in solid lines, in comparison to those in dashed lines in Ref.~\cite{essig}. We set $E_R=1.5\,{\rm keV}$ for comparison.}
  \label{ion}
\end{figure}

The ionization form factor $f_{nl}^{\rm ion}(p_e,q)$ is derived from the bound state out-going wave-functions, obtained for the Schr\"odinger equation with a central potential, as follows,
\begin{equation}
\left|f^{\mathrm{i o n}}_{n l}\left(p_e, q\right)\right|^{2} \equiv \frac{2 p_e^{3}}{(2 \pi)^{3}} \sum_{\mathrm{deg}}\left|f_{n l}(\boldsymbol{q})\right|^{2}
\end{equation}
where  $f_{nl}(\mathbf{q})$ is just the function of radial wave-functions, $\chi_{nl}(k)$, given by
\begin{equation}
\begin{aligned}
\sum_{\mathrm{deg}}\left|f_{n l}(\boldsymbol{q})\right|^{2}
&=(2 l+1) \int_{\left|p_e-q\right|}^{\left|p_e+q\right|} \frac{k d k}{k_2 q}\left|\chi_{n l}(k)\right|^{2}.
\end{aligned}
\end{equation}
 Here,  the  radial wave-functions, $\chi_{nl}(k)$, are expressed in terms of hypergeometric function, 
\bea
\chi_{n l}(p)&=&\sum_{k} C_{n l k} \, 2^{n_{l k}-l}\left(\frac{2 \pi a_{0}}{Z_{l k}}\right)^{3 / 2}\left(\frac{i p \,a_{0}}{Z_{l k}}\right)^{l} \frac{\Gamma\left(n_{l k}+l+2\right)}{\Gamma\left(l+\frac{3}{2}\right) \sqrt{\left(2 n_{l k}\right) !}} \nonumber \\
&&\times\,\,{ }_{2} F_{1}\left[\frac{1}{2}\left(n_{l k}+l+2\right), \frac{1}{2}\left(n_{l k}+l+3\right), l+\frac{3}{2},-\left(\frac{p\, a_{0}}{Z_{l k}}\right)^{2}\right].
\eea
We should mention that the derived form factor cannot reproduce  the atomic response correctly in the regime with low momentum transfer where the dipole approximation holds. So, we set the reference momentum $q_0$ to $20~\mathrm{keV}$ as compared to Ref.~\cite{formfactor},  thus the form factor is modified to 
\begin{equation}
\left|f^{n l}_{\mathrm{ion}}\left(p_{e}, q\right)\right|^{2}=\frac{q^{2}}{q_{0}^{2}} \times\left|f_{n l}^{\mathrm{ion}}\left(p_{e}, q_{0}\right)\right|^{2}
\end{equation}

In Fig.~\ref{event}, we present the ionization form factor $|f_{nl}^{\rm ion}(p_e,q)|^2$ as a function of momentum transfer $q$ in keV units for outer shells ($5p, 4p, 3p$) in Xenon atom in blue, red and green lines, respectively. We have fixed $E_R=1.5\,{\rm keV}$ to compare with the results in the literature shown in dashed lines in the same plot \cite{essig}. Our results for the ionization form factor are shown in solid lines in agreement with those in Ref.~\cite{essig} in dashed lines.

\subsection{Boltzmann equations for dark mesons}

Assuming that the dark matter decoupling takes place at $T_{\pi}>\Delta m$ where $T_\pi$ is the freeze-out temperature of dark mesons, we can take the equal abundances for all the dark mesons by $n_{\tilde{\pi}^0}= n_{\tilde{\pi}^\pm}= \cdots= n_{\tilde{\eta}}\equiv n_{\tilde{\pi}} / N_\pi$.

First, for $m_{Z'}, m_{h'}< m_{\tilde\pi}$, $ \tilde{\pi}\tilde{\pi}\to \tilde{\pi}Z'$, $ \tilde{\pi}\tilde{\pi}\to Z' Z'$ and $ \tilde{\pi}\tilde{\pi}\to h' Z'$ are kinematically open towards zero temperature, so the $2\to 2$ annihilation contributions in the dark sector in the above Boltzmann equation become
\bea
\dot{n}_{\tilde{\pi}} + 3H n_{\tilde{\pi}}&=& -\langle\sigma v^2\rangle_{3\rightarrow 2}(n_{\tilde{\pi}}^3-n_{\tilde{\pi}}^2n_{\tilde{\pi}}^{\rm eq}) - \langle\sigma v\rangle_{2\to 2}(n_{\tilde{\pi}}^2 - (n_{\tilde{\pi}}^{\rm eq})^2) 
\eea
with
\bea
\langle\sigma v\rangle_{2\to 2}=\langle\sigma v\rangle_{\tilde{\pi}\tilde{\pi}\rightarrow e^+e^-}+\langle\sigma v\rangle_{  \tilde{\pi}\tilde{\pi}\to \tilde{\pi}Z'}+\langle\sigma v\rangle_{  \tilde{\pi}\tilde{\pi}\to Z' Z'} +\langle\sigma v\rangle_{  \tilde{\pi}\tilde{\pi}\to h' Z'}+\langle\sigma v\rangle_{  \tilde{\pi}\tilde{\pi}\to h' h'}.
\eea

Second, for $m_{Z'}, m_{h'}> m_{\tilde\pi}$, the Boltzmann equation governing the relic density for dark matter is given by 
\bea
\dot{n}_{\tilde{\pi}} + 3H n_{\tilde{\pi}}&=& -\langle\sigma v^2\rangle_{3\rightarrow 2}(n_{\tilde{\pi}}^3-n_{\tilde{\pi}}^2n_{\tilde{\pi}}^{\rm eq}) - \langle\sigma v\rangle_{\tilde{\pi}\tilde{\pi}\rightarrow e^+e^-}(n_{\tilde{\pi}}^2 - (n_{\tilde{\pi}}^{\rm eq})^2) \nonumber \\
&& + \langle\sigma v\rangle_{\tilde{\pi}Z'\rightarrow \tilde{\pi}\tilde{\pi}}\bigg(n_{\tilde{\pi}} n_{Z'}^{\rm eq}-\frac{n_{Z'}^{\rm eq}}{n_{\tilde{\pi}}^{\rm eq}}n_{\tilde{\pi}}^2\bigg) + \langle\sigma v\rangle_{Z'Z'\rightarrow \tilde{\pi}\tilde{\pi}}\bigg((n_{Z'}^{\rm eq})^2-\frac{(n_{Z'}^{\rm eq})^2}{(n_{\tilde{\pi}}^{\rm eq})^2}n_{\tilde{\pi}}^2\bigg)  \\
&&+ \langle\sigma v\rangle_{h'Z'\rightarrow \tilde{\pi}\tilde{\pi}}\bigg(n_{h'}^{\rm eq} n_{Z'}^{\rm eq} -\frac{n_{h'}^{\rm eq} n_{Z'}^{\rm eq} }{(n_{\tilde{\pi}}^{\rm eq})^2}n_{\tilde{\pi}}^2\bigg)+ \langle\sigma v\rangle_{h'h'\rightarrow \tilde{\pi}\tilde{\pi}}\bigg((n_{h'}^{\rm eq})^2-\frac{(n_{h'}^{\rm eq})^2}{(n_{\tilde{\pi}}^{\rm eq})^2}n_{\tilde{\pi}}^2\bigg). \nonumber 
\eea
Here, the forbidden channels such as $  \tilde{\pi}\tilde{\pi}\to \tilde{\pi}Z'$,   $\tilde{\pi}\tilde{\pi}\to Z' Z'$, $ \tilde{\pi}\tilde{\pi}\to h' Z'$ and $ \tilde{\pi}\tilde{\pi}\to h' h'$ are included in terms of the annihilation cross sections for the inverse processes.

We note that the $3\to 2$ processes and the $2\to 2$ semi-annihilation channels are possible only for $N_f\geq 3$. 
We will also discuss later  the impact of the dark matter self-annihilation on the dark matter freeze-out.

\subsection{Kinetic equilibrium for dark mesons}

For dark matter freeze-out, we assumed that dark matter is in kinetic equilibrium with the SM plasma. Otherwise, the dark matter temperature could differ from the radiation temperature, requiring solving the distribution of dark matter occupancy independently. Moreover, if dark matter annihilation is dominated by $3\to 2$ processes, dark matter keeps getting hot until the low temperature, so it is problematic for the structure formation.

The time evolution of the kinetic energy for dark mesons with $3\to 2$ annihilation processes \cite{vsimp} is dictated by
\bea
{\dot K} +2H K =-m^2_{\widetilde{\pi}} H T^{-1} +T\gamma_\pi(T)
\eea
where $\gamma_\pi(T)$ is the momentum relaxation rate for dark mesons.

From dark meson-electron elastic scattering, ${\tilde\pi}_i e\rightarrow {\tilde\pi}_i e$, we obtain the momentum relaxation rate as
\bea
\gamma_{\tilde\pi}= \frac{40  \zeta(7)}{\pi^3} \frac{ q^2_{{\tilde\pi}_i} \varepsilon^2 e^2 g^2_{Z'}}{ m_{\widetilde{\pi}} m^4_{Z'}}\, T^6 \label{momrelax}
\eea
with  $q_{{\tilde\pi}_i}$ being the dark charge of the dark meson.
Then, the kinetic equilibrium is achieved for $\gamma_\pi(T)>H$ for $2\rightarrow 2$ dominance and  $\gamma_\pi(T)>H (m_{\widetilde{\pi}}/T)^2$ for $3\rightarrow 2$ dominance.
Solely from ${\tilde\pi}_i e\rightarrow {\tilde\pi}_i e$, we can determine the kinetic decoupling temperature as follows:
\bea
T_{\rm kd} =3\,{\rm MeV} \bigg(\frac{g_*}{10.75} \bigg)^{1/12}\bigg(\frac{10^{-4}}{\varepsilon}\bigg)^{1/2}\bigg(\frac{0.6}{q_{{\tilde\pi}_i}g_{Z'}}\bigg)^{1/2} \bigg( \frac{m_{\widetilde{\pi}}}{100\,{\rm MeV}}\bigg)^{1/4}\bigg(\frac{m_{Z'}}{1\,{\rm GeV}} \bigg)
\eea
for $2\rightarrow 2$ dominance, and
\bea
T_{\rm kd} =10\,{\rm MeV} \bigg(\frac{g_*}{10.75} \bigg)^{1/8}\bigg(\frac{10^{-4}}{\varepsilon}\bigg)^{1/3}\bigg(\frac{0.6}{q_{{\tilde\pi}_i}g_{Z'}}\bigg)^{1/3} \bigg( \frac{m_{\widetilde{\pi}}}{100\,{\rm MeV}}\bigg)^{1/2}\bigg(\frac{m_{Z'}}{1\,{\rm GeV}} \bigg)^{2/3}
\label{kdtemp}
\eea
for $3\to 2$ dominance.
But, the kinetic decoupling temperature can be as low as the electron decoupling temperature, due to the decays of $Z'$ or $h'$ into the SM particles.

On the other hand, the scattering between dark mesons and $Z'/h'$, such as ${\tilde\pi}_i Z'\rightarrow {\tilde\pi}_i Z'$ or  ${\tilde\pi}_i h'(Z')\rightarrow {\tilde\pi}_i Z' (h')$ for $m_{h'}\geq m_{Z'} (m_{h'}\leq m_{Z'})$, when accompanied by the decay of $Z'/h'$ into the SM particles, is important for the kinetic equilibrium, as far as $Z'$ and/or $h'$  are in thermal equilibrium with the SM and they have masses comparable or smaller than dark meson masses.
Since the momentum relaxation through the dark scattering with a sizable $g_{Z'}$ is very efficient for $m_{Z'}, m_h' \sim m_{\tilde\pi}$ \cite{vsimp}, the kinetic decoupling temperature is determined by the decay rates of $Z'$ or $h'$, as follows,
\bea
n^{\rm eq}_{Z'} \Gamma_{{Z'\to {\rm SM}}} = H\cdot \Big(n^{\rm eq}_{\tilde\pi}+n^{\rm eq}_{Z'} \Big)
\eea
for  ${\tilde\pi}_i Z'\rightarrow {\tilde\pi}_i Z'$;
\bea
n^{\rm eq}_{Z'(h')} \Gamma_{{Z'(h')\to {\rm SM}}} =H \cdot\Big(n^{\rm eq}_{\tilde\pi}+n^{\rm eq}_{h'(Z')} \Big)
\eea
for  ${\tilde\pi}_i h'(Z')\rightarrow {\tilde\pi}_i Z'(h')$.
Here, the partial decay rates of $Z'$ and $h'$ into an electron-positron pair are given by
\bea
\Gamma(Z'\to e^+e^-) &=&\frac{\varepsilon^2 e^2}{12\pi m_{Z'}} \,(m^2_{Z'}+2m^2_e)\, \bigg(1-\frac{4m^2_e}{m^2_{Z'}} \bigg)^{1/2} , \\
\Gamma(h'\to e^+ e^- ) &=& \frac{m^2_e m_{h'} \sin^2\theta}{8\pi v^2} \,\bigg(1-\frac{4m^2_e}{m^2_{h'}} \bigg)^{3/2}
\eea
where $\theta$ is the mixing angle between the SM and dark Higgs bosons.
We note that if  ${\tilde\pi}_i h'\rightarrow {\tilde\pi}_i h'$ is sizable, it can be also relevant for kinetic equilibrium, but in our model with vector-like quark masses, such channels are suppressed by small mixing Yukawa couplings.

For $m_{Z'}, m_{h'}<m_{\tilde\pi}$, the $Z', h'$ decays in the SM particles are efficient enough such that the kinetic decoupling of dark mesons occurs due to the electron decoupling.
Even for $m_{Z'}, m_{h'}>m_{\tilde\pi}$, as far as $m_{Z'}, m_{h'}$ are comparable to $m_{\tilde\pi}$, the $Z', h'$ decays into the SM particles are efficient enough and the same is true, even with a Boltzmann suppression factor for heavy $Z'$ or $h'$ \cite{vsimp}. 
Therefore, the kinetic decoupling temperature for dark matter in our model is set by the electron decoupling temperature,  $T_e=2m_e$. Then, after the kinetic decoupling of dark matter, the dark matter temperature scales by $T_\chi=\frac{T^2}{T_{\rm kd}}$ with $T$ being the radiation temperature and $T_{\rm kd}=T_e$.

However, we also remark that the kinetic decoupling temperature for dark mesons could be lower than the electron decoupling temperature, if we consider a minimal extension with extra dark particles lighter than $T_e$ such that dark mesons scatters with those particles through $Z'$. In this case, extra light particles can serve as dark radiation to resolve small-scale problems as well as the $H_0$ tension \cite{latedec}.  But, we don't pursue those possibilities further in this work and just show the results for both cases with $T_{\rm kd}=T_e$ and $T_{\rm kd}<T_e$.

\subsection{Late chemical decoupling of dark mesons}

We comment on extra processes for changing the dark matter number such as dark matter decays and annihilations due to mass splittings for dark mesons. 

First, the heavier dark mesons are sufficiently long-lived due to small mass splittings of order a few keV, so the meson decay processes do not determine the dark matter abundance. 

Secondly, quartic self-interactions for dark mesons in the dark chiral perturbation theory lead to the annihilations of heavier dark mesons $\pi_h$ to lighter ones $\pi_l$. Those processes are in equilibrium with the inverse processes until a very low temperature, $T_{\pi}\sim \Delta m$. Thus, as far as the additional $2\rightarrow 2$ annihilation rates are smaller than the Hubble rate at $T_\pi\sim \Delta m$, quartic self-interactions for dark mesons are not relevant for determining the dark matter number density.  
In order for the heavier mesons to be as abundant as the lighter mesons, the $2\rightarrow 2$ annihilation of dark mesons must be decoupled at $T_\pi\gtrsim \Delta m$, that is,  the radiation temperature at the time of the freeze-out must be $T_{\rm f}\gtrsim \sqrt{T_{\rm kd} \Delta m}$ \cite{exodm}. Otherwise, the number densities for the heavier components would be Boltzmann suppressed as $n_{\pi_h}=e^{-\Delta m/T_\pi}\, n_{\pi_l}$.
For  kinetic decoupling temperature $T_{\rm kd}\sim 1\,{\rm MeV} (1\,{\rm keV})$ and $\Delta m=2.5\,{\rm keV}$, we need $T_{\rm f}\gtrsim 50(1.6)\,{\rm keV}$. 

We impose the chemical decoupling condition for  the $2\rightarrow 2$ annihilation, as follows, 
\bea
n_{\pi_h} \langle\sigma v\rangle_{\pi_h \pi_h\rightarrow \pi_l\pi_l} = H \qquad {\rm at}\,\,\, T=T_{\rm f} >\sqrt{ T_{\rm kd}\Delta m} \label{latedec0}
\eea
where  $H=0.33 \, g^{1/2}_* T^2/M_P$ and the annihilation cross sections for $\pi_h\pi_h\to \pi_l\pi_l$  is parametrized by
\bea
\langle\sigma v\rangle_{\pi_h\pi_h\rightarrow \pi_l\pi_l} =\frac{\alpha^2_{\rm eff}}{m^2_{\tilde\pi}}\, \sqrt{\frac{\Delta m}{m_{\widetilde{\pi}}}} \label{dmselfann}
\eea
with $\alpha_{\rm eff}$ being the effective coupling for the annihilation cross section. 
Since the number density of the heavier dark meson at freeze-out is given by
\bea
n_{\pi_h}(T_{\rm f})&=& n_{\pi_h}(T_0)\,\cdot \bigg(\frac{g_{*s}(T_{\rm f})T^3_{\rm f} }{g_{*s}(T_0)T^3_0}\bigg) \nonumber \\
&=&\frac{\Omega_{\pi_h}}{m_{\tilde\pi}}\,\cdot \rho_c(T_0)\, \cdot \bigg(\frac{g_{*s}(T_{\rm f})T^3_{\rm f} }{g_{*s}(T_0)T^3_0}\bigg)
\eea
where $\Omega_{\pi_h}$ is the density fraction of the heavier component at present and $\rho_c(T_0)=8.1\times 10^{-47}\,h^2\,{\rm GeV}^4$ is the critical density at present.
Consequently,  the decoupling condition  in eq.~(\ref{latedec0}) determines the radiation temperature at the time of freeze-out to be
\bea
T_{\rm f} = \bigg(\frac{0.24\,{\rm GeV}^{-2}}{\alpha^2_{\rm eff}/m^2_{\tilde\pi}}\bigg) \,\bigg(\frac{0.12/8}{\Omega_{\pi_h} h^2}\bigg) \bigg(\frac{g_*(T_{\rm f})}{3.36}\bigg)^{1/2}\bigg(\frac{3.91}{g_{*s}(T_{\rm f})}\bigg)\bigg(\frac{m_{\widetilde{\pi}}/\Delta m}{2\times 10^5}\bigg)^{3/2}\Delta m,
\eea
resulting with $T_{\rm f} >\sqrt{ T_{\rm kd}\Delta m} $ in the following upper bound on the effective coupling,
\bea
\frac{\alpha^2_{\rm eff}}{m^2_{\tilde\pi}} \lesssim 0.24\,{\rm GeV}^{-2}\,\bigg(\frac{0.12/8}{\Omega_{\pi_h} h^2}\bigg) \bigg(\frac{g_*(T_{\rm f})}{3.36}\bigg)^{1/2}\bigg(\frac{3.91}{g_{*s}(T_{\rm f})}\bigg)\bigg(\frac{m_{\widetilde{\pi}}/\Delta m}{2\times 10^5}\bigg)^{3/2} \bigg(\frac{\Delta m}{T_{\rm kd}}\bigg)^{1/2}. \label{latedec}
\eea
If the above bound is not satisfied, the relic densities of heavier dark mesons would get suppressed by $\Omega_{\pi_h}=e^{-\Delta m T_{\rm kd}/T^2_{\rm f}}\, \Omega_{\pi_l}$, as compared to those for lighter dark mesons.

\subsection{Lifetime of heavy dark mesons}

The $Z'$ portal interactions allow the heavier meson to decay into the lighter one plus a neutrino pair, with the decay rate \cite{exodm},
\bea
\Gamma({\widetilde \pi}_i\rightarrow {\widetilde \pi}_j\nu{\bar \nu})
&\simeq& \frac{ N_\nu q^2_{{\tilde\pi}_i} e^2 \varepsilon^2 g^2_{Z'}(\Delta m_{ij})^5}{1920 \pi^3 c_W^4 m^4_Z} \nonumber\\
&\simeq & (2.7\times 10^{26}\,{\rm sec})^{-1} N_\nu \bigg(\frac{\varepsilon}{10^{-4}} \bigg)^2 \bigg(\frac{q_{{\tilde\pi}_i} g_{Z'}}{0.2} \bigg)^2 \bigg(\frac{\Delta m_{ij}}{3\,{\rm keV}} \bigg)^5
\eea
with $q_{{\tilde\pi}_i}$ being the dark charge of the dark meson.
On the other hand, there is no decay mode with two photons, ${\widetilde \pi}_i\rightarrow {\widetilde \pi}_j\gamma\gamma$, etc, due to the absence of the effective coupling for $Z'-\gamma-\gamma$ \cite{exodm}. The three-photon decay channels can be open at loops, but they are sufficiently suppressed to be consistent with the $X$-ray bounds.
Therefore, all the dark mesons can be sufficiently long-lived to make up for dark matter in the Universe at present, as far as the late decoupling condition is satisfied as discussed in eq.~(\ref{latedec}).

\section{Benchmark models for XENON1T excess and constraints}

We make a concrete discussion on the exothermic processes for explaining the XENON1T excess in models for dark mesons with $N_f=3$. We apply various constraints on the model discussed in the previous section and impose experimental bounds on the parameter space that is compatible the XENON1T excess.

\subsection{Exothermic process for XENON1T}

\begin{figure}[tbp]
  \centering
\includegraphics[width=.60\textwidth]{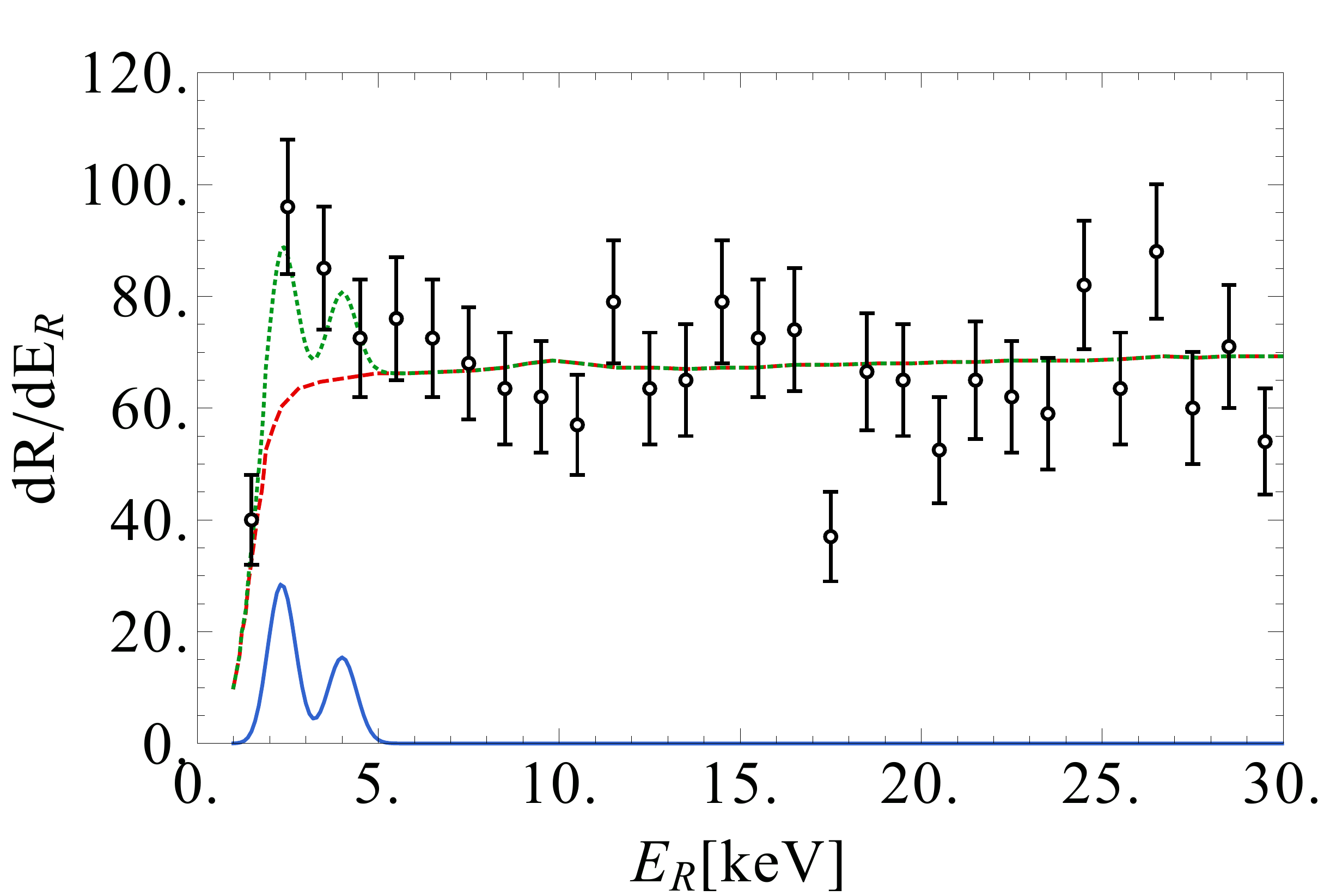}
  \caption{Event rate for electron recoil as a function of recoil energy $E_R$ in keV. Blue line indicates the signal events from exothermic processes of dark mesons and green line shows the combined signal and background events.  The red line is the background model used by Xenon experiment \cite{xenon}. We have taken $\Delta m=4.0\,{\rm keV}$ and ${\bar\sigma}_e/m_{\chi_1}=1.2\times 10^{-43}\,{\rm cm}^2/{\rm GeV}$. }
  \label{event}
\end{figure}

For $N_f=3$, assuming that the $Z'$ corrections to dark meson masses  are ignored, we have shown in the previous section that there is a mass hiearchy, $m_{{\widetilde\eta}^0}>m_{{\widetilde K}^0}>m_{{\widetilde\pi}^\pm}>m_{{\widetilde K}^\pm}>m_{{\widetilde\pi}^0}$, there are multiple exothermic scattering processes between dark mesons and electron: ${\widetilde K}^0 e\rightarrow {\widetilde K}^+ e$, $\overline{{\widetilde K}^0} e\rightarrow {\widetilde K}^- e$, and ${\widetilde\pi}^\pm e\rightarrow {\widetilde\pi}^0 e$. 
Then, for instance, taking $\Delta m/\sqrt{3}=2.5\,{\rm keV}$ from  ${\widetilde\pi}^\pm e\rightarrow {\widetilde\pi}^0 e$, we have a monochromatic electron recoil energy at $E_R\simeq \Delta m/\sqrt{3}=2.5\,{\rm keV}$ and there appears another peak at $\Delta m=4.3\,{\rm keV}$ from ${\widetilde K}^0 e\rightarrow {\widetilde K}^+ e$.
As a result, we get the total event rate per Xenon detector for dark mesons  as
\bea
R_D 
&\simeq &50 \bigg(\frac{M_T}{\rm tonne-yrs}\bigg) \bigg(\frac{2K_{\rm int}(\Delta m)\cdot{\rho_{{\widetilde K}^0}+\frac{4}{3^{1/4}}\,K_{\rm int}\big(\frac{1}{\sqrt{3}}\Delta m\big)\cdot \rho_{{\widetilde\pi}^+}}}{2.6\cdot(0.4\,{\rm GeV\, cm^{-3}})} \bigg)  \nonumber \\
&&\times \bigg(\frac{{\bar\sigma}_e/m_{\pi}}{1.2\times 10^{-43}\,{\rm cm}^2/{\rm GeV}} \bigg) \bigg(\frac{\Delta m}{2.5\,{\rm keV}}\bigg)^{1/2} \label{totaln3}
\eea
For $K_{\rm int}(\Delta m)\simeq K_{\rm int}\big(\frac{1}{\sqrt{3}}\Delta m\big)$,  the event rate at $E_R\simeq \Delta m/\sqrt{3}$ due to  ${\widetilde\pi}^\pm e\rightarrow {\widetilde\pi}^0 e$ is about twice the even rate at $E_R\simeq \Delta m$ due to  ${\widetilde K}^0 e\rightarrow {\widetilde K}^+ e$.

In Fig.~\ref{event}, we show the event rate for electron recoil as a function of recoil energy $E_R$ in keV on the right of Fig.~\ref{event}. The background model taken by Xenon experiment \cite{xenon} is shown in red dashed line, and the portion in blue is the signal from exothermic dark mesons in our model, and finally the dashed green line is the sum of the background and signal events, in comparison to the XENON events with black bars. We set $\Delta m/\sqrt{3}=2.3\,{\rm keV}$ (so $\Delta m=4.0\,{\rm keV}$) and ${\bar\sigma}_e/m_{\chi_1}=1.2\times 10^{-43}\,{\rm cm}^2/{\rm GeV}$, so there are two monochromatic peaks at $E_R=2.3\,{\rm keV}$ and $4.0\,{\rm keV}$ in the electron recoil energy spectrum.
Although the two peaks are smeared out after being convoluted with the detector resolution \cite{xenon,exodm}, the difference between them is resolvable with the detector resolution and it could be distinguishable from the case with a single peak.

\begin{figure}[tbp]
  \centering
\includegraphics[width=.48\textwidth]{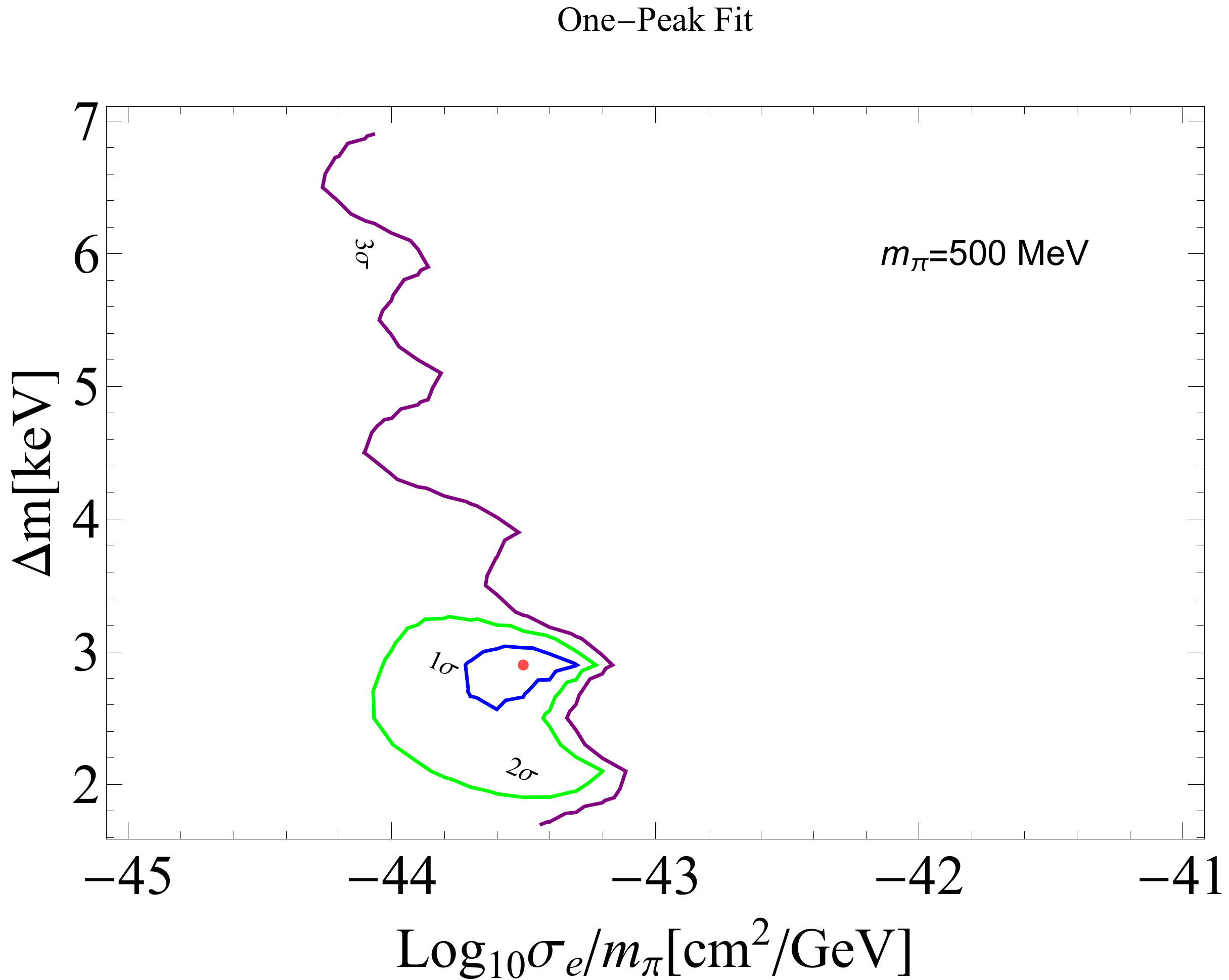}\,\,
\includegraphics[width=.48\textwidth]{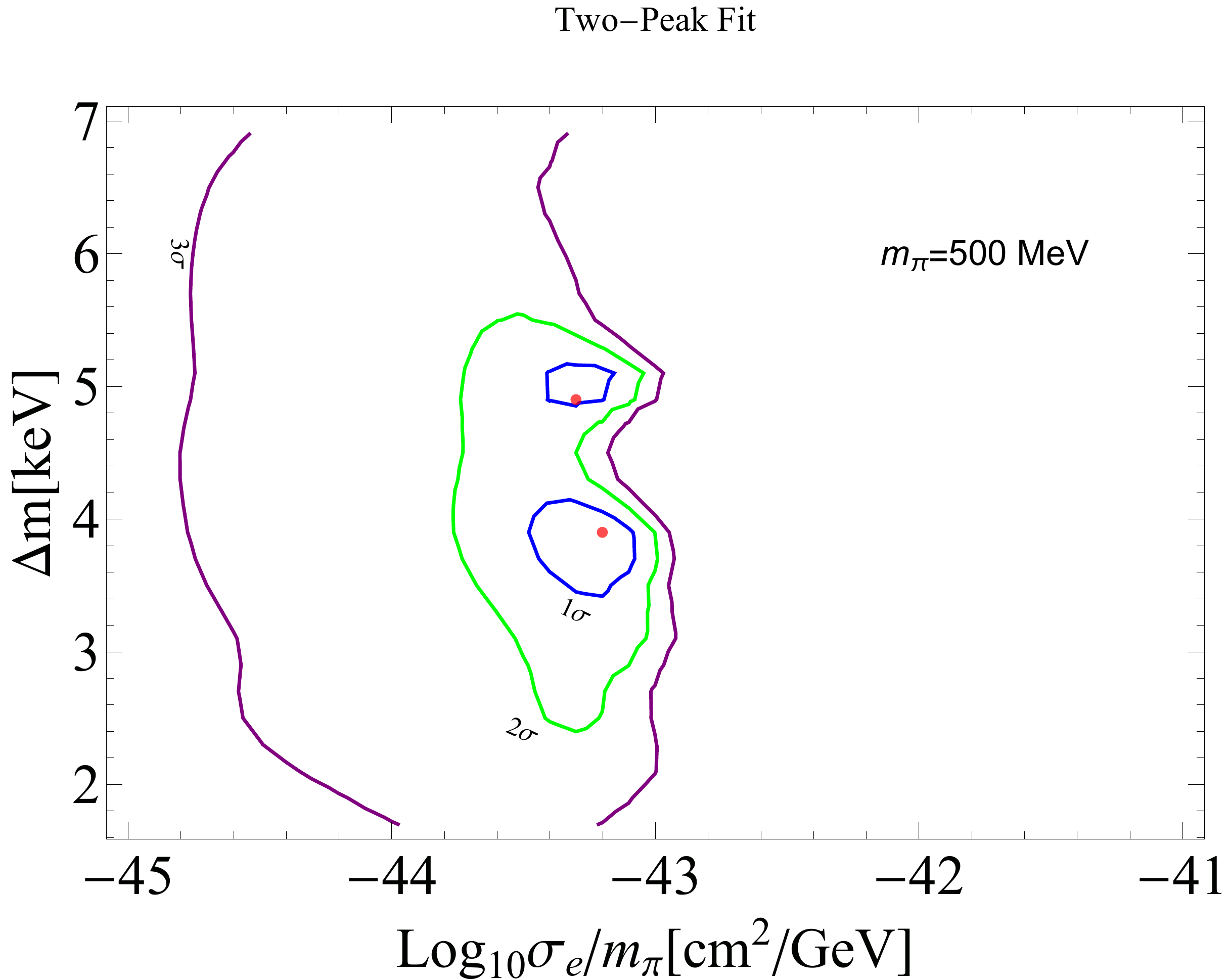}
  \caption{The minimum $\chi^2$ fit to XENON1T data in the parameter space for $\Delta m$ (in keV) vs ${\bar \sigma}_e/m_{\tilde\pi}$ (in ${\rm cm^2/GeV}$), for the case with one peak on left and the case with two peaks on right. Blue, green and purple lines indicate $1\sigma, 2\sigma$ and $3\sigma$ contours from the minimum $\chi^2$, respectively. The best-fit points with local minimum $\chi^2$ is shown in red. Dark meson masses are fixed to $m_{\tilde\pi}=500\,{\rm MeV}$, but the results are insensitive to other values as far as $m_{\tilde\pi}>10\,{\rm MeV}$.  }
  \label{bestfit}
\end{figure}

In Fig.~\ref{bestfit}, we also depict the minimum $\chi^2$ fit of the parameter space in $\Delta m$ in keV vs ${\bar \sigma}_e/m_{\tilde\pi}$ in ${\rm cm^2/GeV}$ to electron recoil energy in XENON1T. The plot on right in Fig.~\ref{bestfit} is the fit result with two monochromatic peaks from exothermic dark matter in electron recoil energy in our work, and the plot on left in Fig.~\ref{bestfit} is also shown for comparison to the case with one monochromatic peak from exothermic dark matter, as discussed in Ref.~\cite{exodm}. We fixed common dark meson masses to $m_{\tilde\pi}=500\,{\rm MeV}$, but the results are insensitive to other values of dark meson masses as far as $m_{\tilde\pi}>10\,{\rm MeV}$. 

The global best fit values for the two-peak case are 
\bea
[\Delta m=3.9\,{\rm keV},\,\, {\bar\sigma}_e/m_{\tilde\pi}= 2\times10^{-43}\,{\rm cm^2/GeV}],
\eea
with $\chi^2_{\rm min}=1.48$, 
and there is another local best-fit point for $[\Delta m=4.9\,{\rm keV},{\bar\sigma}_e/m_{\tilde\pi}= 2\times10^{-43}\,{\rm cm^2/GeV}]$, with $\chi^2_{\rm min}=3.64$.
In comparison, the best fit values for the one-peak case are $[\Delta m=2.9\,{\rm keV}, {\bar\sigma}_e/m_{\tilde\pi}= 1.3\times10^{-43}\,{\rm cm^2/GeV}]$, with $\chi^2_{\rm min}= 2.43$. 
Therefore, the two-peak case in our work shows a slightly better fit to XENON1T data as compared to the one-peak case.

\subsection{Dark matter annihilations}

The $3\rightarrow 2$ annihilation with WZW terms can be important for the freeze-out process for dark mesons with degenerate masses in the strongly coupled regime \cite{simpmeson1,simpmeson2,simpmeson3}. 
In this case, the typical dark meson masses with a correct relic density are about $100\,{\rm MeV}$.
The corresponding annihilation cross section for $3\rightarrow 2$ processes is given by
\bea
\langle\sigma v^2\rangle_{3\rightarrow 2} = \frac{5\sqrt{5} N^2_c m^5_\pi}{2048\pi^5 f^{10}_\pi} \frac{t^2}{N^3_\pi} \, x^{-2}
\eea
where $x\equiv m_{\widetilde{\pi}}/T$,  $N_\pi$ is the number of dark mesons and $t^2$ is the group theory factor, given by $t^2=\frac{4}{3} N_f (N^2_f-1)(N^2_f-4)$ for $SU(N_f)$ flavor symmetry \cite{simpmeson1}. In our case, we took $N_\pi=8$ and $N_f=3$, for which $t^2=160$.

Moreover, dark mesons charged under $Z'$ can annihilate into a pair of light charged particles in the SM through $Z'$ mediator, that is, by ${\widetilde K}^+{\widetilde K}^-\rightarrow e^+ e^-$, ${\widetilde K}^+ {\widetilde K}^0\rightarrow e^+ e^-$, ${\widetilde K}^0 \overline{{\widetilde K}^0}\rightarrow e^+ e^-$, ${\widetilde K}^- \overline{{\widetilde K}^0}\rightarrow e^+ e^-$, and ${\widetilde \pi}^0{\widetilde \pi}^\pm\rightarrow e^+ e^-$, thus, if  the corresponding $2\rightarrow 2$ annihilation cross section is sizable,  the standard freeze-out is achieved at a higher temperature.
In this case, the $2\rightarrow 2$ annihilation cross sections are given by
\bea
\langle\sigma v\rangle_{{\widetilde K}^+{\widetilde K}^-\rightarrow e^+ e^-} 
&=&\langle\sigma v\rangle_{{\widetilde K}^+ \overline{{\widetilde K}^0}\rightarrow e^+ e^-}= \langle\sigma v\rangle_{{\widetilde K}^0 \overline{{\widetilde K}^0}\rightarrow e^+ e^-}\nonumber \\
&=& \langle\sigma v\rangle_{{\widetilde K}^-{\widetilde K}^0\rightarrow e^+ e^-}
=\frac{1}{2} \langle\sigma v\rangle_{{\widetilde \pi}^0{\widetilde \pi}^\pm\rightarrow e^+ e^-}  \nonumber \\
&=&\frac{\varepsilon^2 e^2 g^2_{Z'}}{\pi} \frac{m^2_{\widetilde{\pi}} + \frac{1}{2} m^2_e}{(4m^2_{\widetilde{\pi}}-m^2_{Z'})^2+\Gamma^2_{Z'} m^2_{Z'}}\, \sqrt{1-\frac{m^2_e}{m^2_{\widetilde{\pi}}}}\, x^{-1}\,.
\eea
Then, the effective $2\to 2$ annihilation cross section into the visible sector is given by 
\bea
\langle\sigma v\rangle_{\tilde{\pi}\tilde{\pi}\rightarrow e^+e^-}= \frac{16}{N^2_\pi} \,\langle\sigma v\rangle_{{\widetilde K}^+{\widetilde K}^-\rightarrow e^+ e^-}. 
\eea

If $m_{Z'}, m_{h'}< m_{\widetilde{\pi}}$,  there are additional $2\to 2$ annihilation channels, ${\widetilde\pi}^i{\widetilde\pi}^j\rightarrow {\widetilde\pi}^j Z'$,  ${\widetilde\pi}^i{\widetilde\pi}^j\rightarrow Z'Z'$ as well as ${\widetilde\pi}^i{\widetilde\pi}^j\rightarrow h' Z'$. In this case, the corresponding $2\to 2$ annihilation  cross sections are given by
\bea
\langle \sigma v \rangle_{ \tilde{\pi}\tilde{\pi}\to\tilde{\pi} Z' } &=&\frac{g_{Z'}^2N_c^2}{576\pi^5 m_\pi^2 N_\pi^2}\bigg(\frac{m_\pi}{f_\pi}\bigg)^6\bigg(1-\frac{m_{Z'}^2}{m_\pi^2}\bigg)^{3/2}\bigg(9-\frac{m_{Z'}^2}{m_\pi^2}\bigg)^{3/2}x^{-1}, \\
\langle \sigma v \rangle_{  \tilde{\pi}\tilde{\pi} \to Z' Z'} &=& \frac{4g_{Z'}^4m_\pi^2}{\pi (2m_\pi^2-m_{Z'}^2)^2 N_\pi^2}\bigg(8-8\frac{m_Z'^2}{m_\pi^2}+3\frac{m_{Z'}^4}{m_\pi^4}\bigg)\sqrt{1-\frac{m_{Z'}^2}{m_\pi^2}}, \\
\langle \sigma v \rangle_{  \tilde{\pi}\tilde{\pi} \to h' Z'} &=&\frac{g_{Z'}^4m_\pi^2}{\pi (4m_\pi^2-m_{Z'}^2)^2N_\pi^2}\Big(16-8\frac{m_{h'}^2}{m_\pi^2}+40\frac{m_{Z'}^2}{m_\pi^2}+\frac{m_{h'}^4}{m_\pi^4}-2\frac{m_{h'}^2m_{Z'}^2}{m_{\pi}^4}+\frac{m_{Z'}^4}{m_{\pi}^4}\Big) \nonumber \\
&&\quad \times \Big(1-\frac{(m_{h'}-m_{Z'})^2}{4m_\pi^2}\Big)^{1/2}\Big(1-\frac{(m_{h'}+m_{Z'})^2}{4m_\pi^2}\Big)^{1/2}x^{-1}.
\eea
On the other hand, the $ \tilde{\pi}\tilde{\pi}\to h' h'$ channels are suppressed, because dark quarks are vector-like so there is no diagonal Yukawa coupling of dark quarks to the dark Higgs.

Next, for $m_{Z'}\gtrsim m_{\widetilde{\pi}}$, the additional $2\rightarrow 2$ annihilation channels are Boltzmann-suppressed, but they can be relevant for light dark matter if the $Z'$ mass is close to dark meson masses.
For $m_{Z'}\gg m_{\widetilde{\pi}}$, the forbidden  $2\rightarrow 2$ annihilation cross sections can be sufficiently suppressed.
The effective annihilation cross sections for the forbidden channels are given by
\bea
\langle \sigma v \rangle_{ \tilde{\pi}\tilde{\pi}\to\tilde{\pi} Z' } &=&\bigg(\frac{n^{\rm eq}_{Z'}}{n^{\rm eq}_{\tilde{\pi}}}\bigg) \langle \sigma v \rangle_{\tilde{\pi} Z'\rightarrow \tilde{\pi}\tilde{\pi}}, \\
\langle \sigma v \rangle_{  \tilde{\pi}\tilde{\pi} \to Z' Z'} &=&\bigg(\frac{n^{\rm eq}_{Z'}}{n^{\rm eq}_{\tilde{\pi}}}\bigg)^2  \langle \sigma v \rangle_{Z' Z'\rightarrow \tilde{\pi}\tilde{\pi}}, \\
\langle \sigma v \rangle_{  \tilde{\pi}\tilde{\pi} \to h' Z'} &=&\bigg(\frac{n^{\rm eq}_{Z'}n^{\rm eq}_{h'}}{(n^{\rm eq}_{\tilde{\pi}})^2}\bigg)\langle \sigma v \rangle_{h' Z' \to\tilde{\pi}\tilde{\pi} } 
\eea
with
\bea
\langle \sigma v \rangle_{\tilde{\pi} Z'\rightarrow \tilde{\pi}\tilde{\pi}} &= & \frac{g_{Z'}^2 N_c^2 m_{\widetilde{\pi}} m_{Z'}^4}{108\pi^5 N_\pi f_\pi^6 (m_{\widetilde{\pi}}+m_{Z'})}\bigg(1-\frac{m_{\widetilde{\pi}}}{m_{Z'}}\bigg)^{3/2}\bigg(1+3\frac{m_{\widetilde{\pi}}}{m_{Z'}}\bigg)^{3/2}x^{-1},   \\
\langle \sigma v \rangle_{Z' Z'\rightarrow \tilde{\pi}\tilde{\pi}} &=& \frac{4g_{Z'}^4}{9\pi m_{Z'}^2}\bigg( 11-24\frac{m_\pi^2}{m_{Z'}^2} + 16\frac{m_\pi^4}{m_{Z'}^4} \bigg)\sqrt{1-\frac{m_\pi^2}{m_{Z'}^2}}, \\
\langle \sigma v \rangle_{h' Z' \to \tilde{\pi}\tilde{\pi} } &=&\frac{8 g_{Z'}^4 m_{Z'}}{3m_{h'}^3\pi}\Big(\frac{m_{h'}+m_{Z'}}{m_{h'}+2m_{Z'}}\Big)^2\Big(1-\frac{4m_\pi^2}{(m_{h'}+m_{Z'})^2}\Big)^{3/2}.
\eea
We note that the $ \tilde{\pi}\tilde{\pi}\to h' h'$ forbidden channels are further suppressed due to small mixing Yukawa couplings between dark quarks and dark Higgs in our model, but they are omitted.

\subsection{Dark meson self-annihilations}

First, we note that for $N_f=3$, dark mesons can be in kinetic equilibrium through ${\widetilde K}^0 e\rightarrow {\widetilde K}^0 e$, ${\widetilde K}^0 e\rightarrow {\widetilde K}^+ e$, and  ${\widetilde \pi}^\pm e\rightarrow {\widetilde \pi}^0 e$, each of which has the same momentum relaxation rate given by eq.~(\ref{momrelax}) with $q_{{\tilde \pi}_i}=2$. In this case, the kinetic decoupling temperature for dark mesons would be about $T_{\rm kd}\sim 3\,{\rm MeV}$, as discussed in Section 4.3.

As we discussed in Section 4. 3, however, since the scattering between dark mesons and $Z'$ (or $h'$) with comparable masses to dark mesons is efficient, the small decay fractions of $Z'$ or $h'$ into the SM particles are sufficient to maintain the kinetic equilibrium until the kinetic decoupling temperature of electron, which is $T_{\rm kd}=1\,{\rm MeV}$. 
Furthermore, the scattering between dark mesons and extra light fermions in the dark sector \cite{latedec} could delay the kinetic decoupling to as low as $T_{\rm kd}=1\,{\rm keV}$ without changing the dark matter freeze-out, so we keep the kinetic decoupling temperature to be a variable parameter in our model. 

There are $2\rightarrow 2$ self-annihilations of dark mesons such as ${\widetilde K}^0 \overline{{\widetilde K}^0}\rightarrow {\widetilde K}^+{\widetilde K}^-$, ${\widetilde\pi}^\pm {\widetilde\pi}^\mp\to {\widetilde\pi}^0 {\widetilde\pi}^0$, etc.
The relevant self-interactions for the $2\rightarrow 2$ annihilation of heavier dark mesons are 
\bea
{\cal L}_{4{\widetilde\pi}}&\supset& \frac{1}{3f^2_\pi} \Big[{\widetilde\pi}^0{\widetilde\pi}^+(\partial_\mu{\widetilde\pi}^0\partial^\mu {\widetilde\pi}^-)+{\widetilde\pi}^0{\widetilde\pi}^-(\partial_\mu{\widetilde\pi}^0\partial^\mu {\widetilde\pi}^+)  
-{\widetilde\pi}^+ {\widetilde\pi}^- (\partial_\mu{\widetilde\pi}^0 )^2-{\widetilde\pi}^0 {\widetilde\pi}^0 (\partial_\mu{\widetilde\pi}^+ \partial^\mu{\widetilde\pi}^-)
\nonumber \\
&&\quad -\frac{1}{2} {\widetilde K}^+{\widetilde K}^-(\partial_\mu{\widetilde K}^0\partial^\mu \overline{{\widetilde K}^0})
-\frac{1}{2}{{\widetilde K}^0}\overline{{\widetilde K}^0}(\partial_\mu{{\widetilde K}^+}\partial^\mu {{\widetilde K}^-})
-\frac{1}{2} \overline{{\widetilde K}^0}{\widetilde K}^+(\partial_\mu{\widetilde K}^0\partial^\mu{{\widetilde K}^-})
 \nonumber \\
&&\quad
-\frac{1}{2} {{\widetilde K}^0}{\widetilde K}^-(\partial_\mu\overline{{\widetilde K}^0}\partial^\mu {{\widetilde K}^+})
+{{\widetilde K}^0}{\widetilde K}^+(\partial_\mu\overline{{\widetilde K}^0}\partial^\mu {{\widetilde K}^-})
+\overline{{\widetilde K}^0}{\widetilde K}^-(\partial_\mu{\widetilde K}^0\partial^\mu {{\widetilde K}^+}) \Big].
\eea
Then, the effective couplings for the  annihilation cross sections for $\pi_h \pi_h\to \pi_l \pi_l $ in the parameterization in eq.~(\ref{dmselfann}) are given by
\bea
\alpha^2_{\rm eff}=\left\{ \begin{array}{c}   \frac{m_{\widetilde{\pi}}^4}{288 \sqrt{2} \times 3^{1/4}\pi f_\pi^4}
\bigg(7+48\frac{f_\pi^2 g_{Z'}^2}{m_{Z'}^2}\bigg)^2,\qquad  {\widetilde\pi}^\pm {\widetilde\pi}^\mp\to {\widetilde\pi}^0 {\widetilde\pi}^0, \vspace{0.2cm} \\   \frac{m_{\widetilde{\pi}}^4}{144 \sqrt{2} \pi f_\pi^4}\bigg(1+12\frac{f_\pi^2 g_{Z'}^2}{m_{Z'}^2}\bigg)^2, \qquad{\widetilde K}^0 \overline{{\widetilde K}^0}\rightarrow {\widetilde K}^+{\widetilde K}^-.  \end{array} \right. \label{alpeff}
\eea
For $g_{*s}(T_f)=3.91$ and $g_{*}(T_f)=3.36$, from eq.~(\ref{latedec}), we find  the late decoupling conditions on the self-interactions of dark mesons, as follows,
\bea
\frac{m_{\widetilde{\pi}}}{f_\pi}\bigg(1+\frac{48}{7}\frac{f_\pi^2 g_{Z'}^2}{m_{Z'}^2}\bigg)^{1/2} \lesssim   0.12\, \Big(\frac{m_{\widetilde{\pi}}}{100\,{\rm MeV}}\Big)^{1/2}\bigg(\frac{0.12/8}{\Omega_h h^2}\bigg)^{1/4}  \bigg(\frac{m_{\widetilde{\pi}}/\Delta m}{2.5\times 10^4}\bigg)^{3/8}\bigg(\frac{\Delta m/4\,{\rm keV}}{T_{\rm kd}/1\,{\rm MeV}}\bigg)^{1/8}  \label{latedec1}
\eea
for $ {\widetilde\pi}^\pm {\widetilde\pi}^\mp\to {\widetilde\pi}^0 {\widetilde\pi}^0$,
and
\bea
\frac{m_{\widetilde{\pi}}}{f_\pi} \bigg(1+12\frac{f_\pi^2 g_{Z'}^2}{m_{Z'}^2}\bigg)^{1/2}\lesssim 0.26\, \Big(\frac{m_{\widetilde{\pi}}}{100\,{\rm MeV}}\Big)^{1/2}\bigg(\frac{0.12/8}{\Omega_h h^2}\bigg)^{1/4}  \bigg(\frac{m_{\widetilde{\pi}}/\Delta m}{2.5\times 10^4}\bigg)^{3/8}\bigg(\frac{\Delta m/4\,{\rm keV}}{T_{\rm kd}/1\,{\rm MeV}}\bigg)^{1/8}  \label{latedec2}
\eea
for $ {\widetilde K}^0 \overline{{\widetilde K}^0}\rightarrow {\widetilde K}^+{\widetilde K}^-$.
Therefore, the chiral perturbation theory for dark mesons is in the weakly coupled regime with $m_{\tilde\pi}/f_\pi\lesssim 1$ to be compatible with the XENON1T excess, so the $3\to 2$ annihilation processes are subdominant for determining the relic density for dark mesons. 

We remark that the self-interactions of dark mesons can be as large as the perturbativity bound, $m_{\tilde\pi}/f_\pi= 2\pi$, being consistent with the late decoupling, for $m_{\tilde\pi}=100\,{\rm MeV}$ and $T_{\rm kd}=1\,{\rm MeV}$, provided that $\Delta m\lesssim 0.3-0.5\,{\rm keV}$.  Then, the $3\to 2$ annihilation processes can be dominant for the relic density of dark mesons \cite{simpmeson1,simpmeson2,simpmeson3}. But, in this case, we could not explain the Xenon excess.

\begin{figure}[tbp]
  \centering
\includegraphics[width=.45\textwidth]{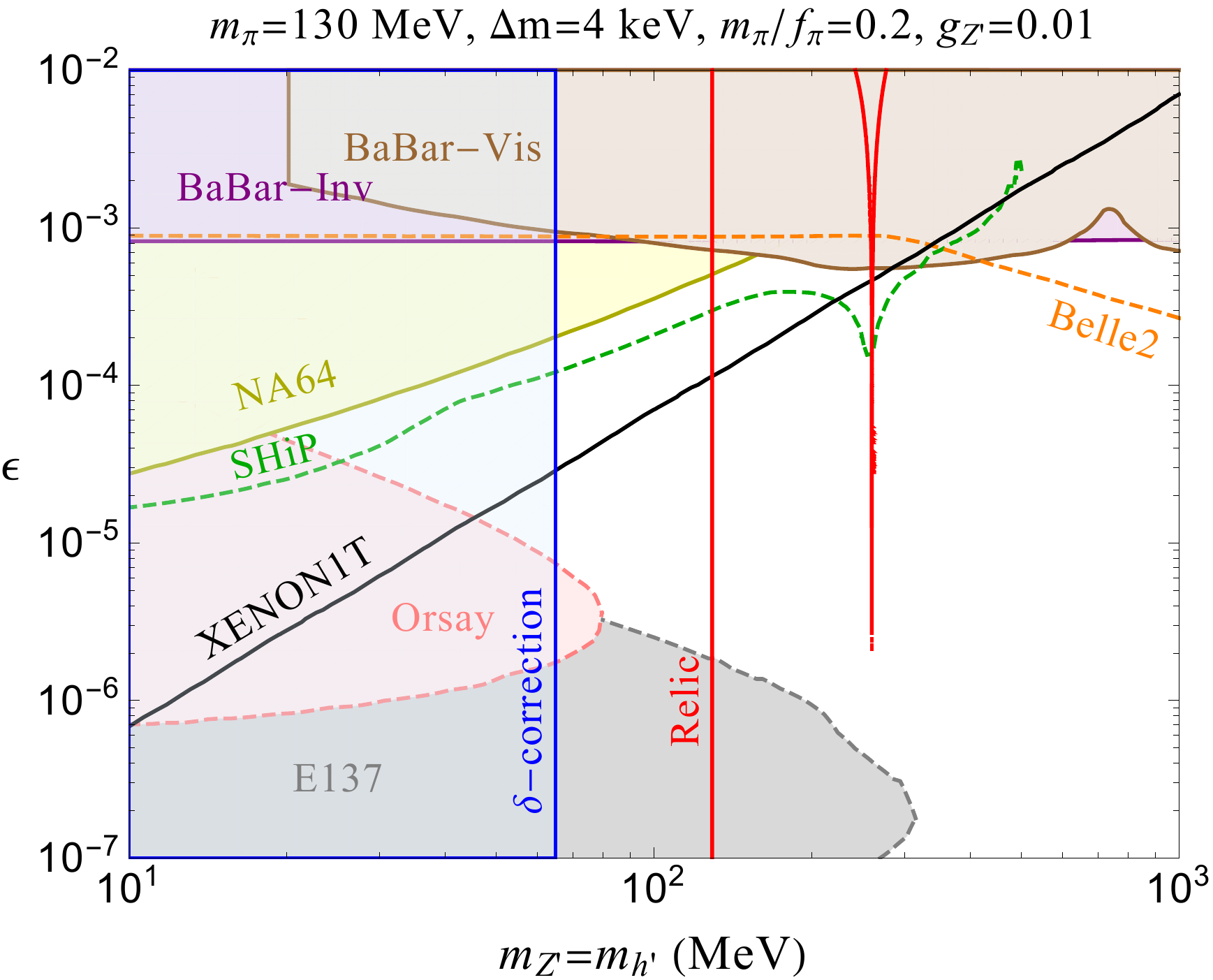}\quad
\includegraphics[width=.45\textwidth]{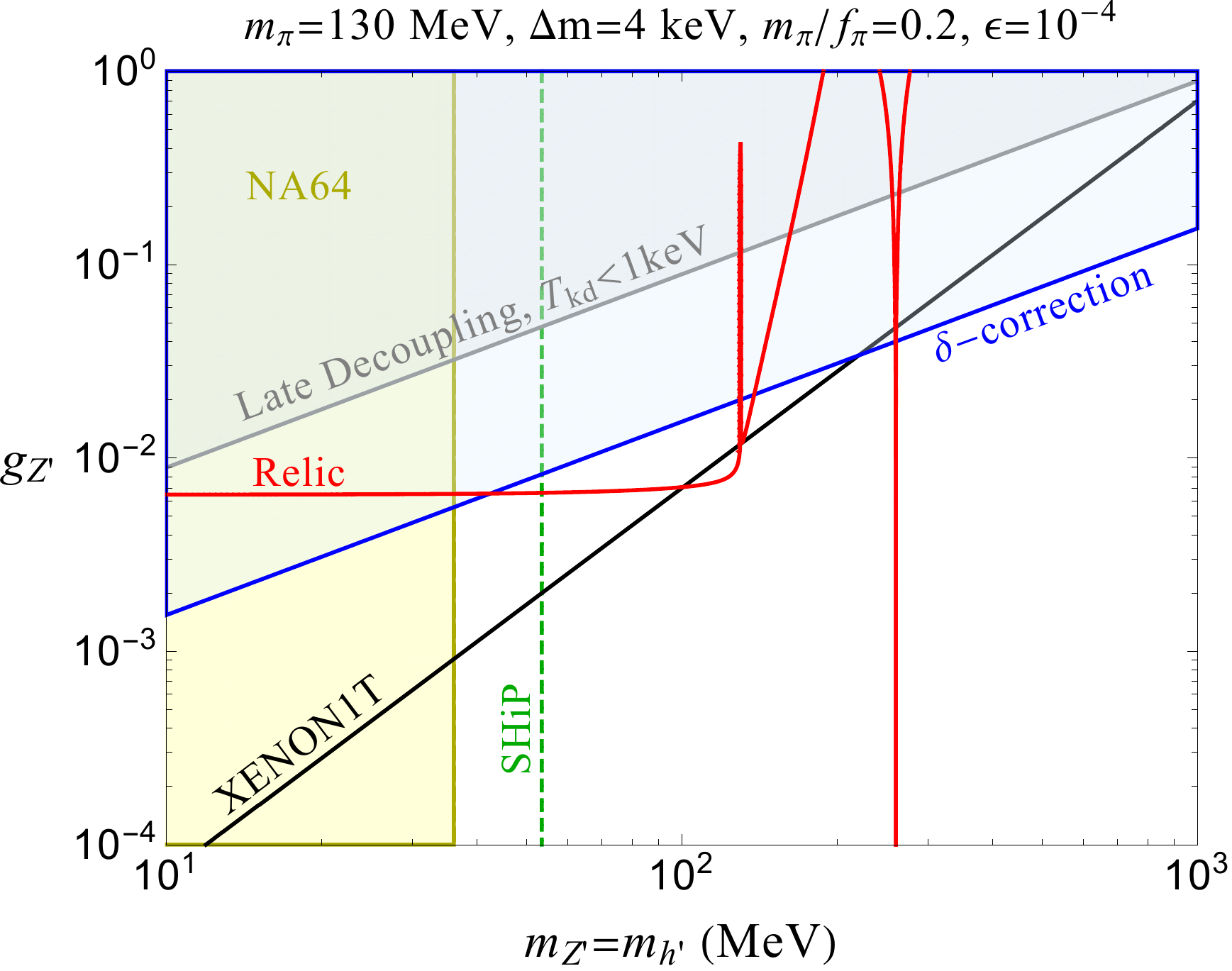}  \vspace{0.5cm} \\
\includegraphics[width=.45\textwidth]{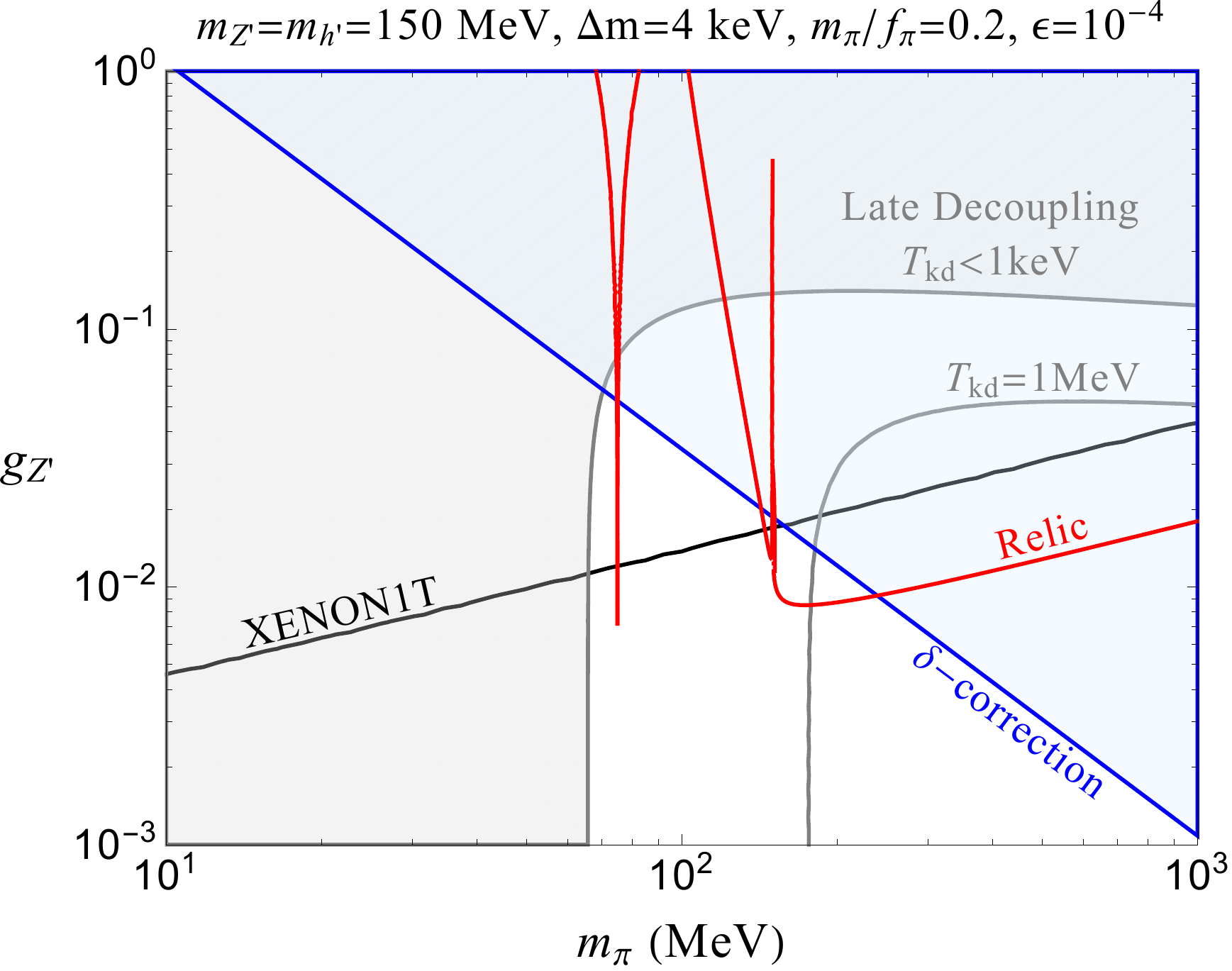} 
  \caption{Parameter space for explaining the XENON1T electron excess: $\varepsilon$ vs $m_{Z'}$ (left in top), $g_{Z'}$ vs $m_{\tilde\pi}$ (right in top) and $g_{Z'}$ vs $m_{\tilde\pi}$ (bottom). The XENON1T electron excess can be explained along black line and the correct relic density is saturated along red line. The relic densities for heavier dark mesons are exponentially suppressed in gray region (line) due to late decoupling for $T_{\rm kd}<1\,{\rm keV}\, (T_{\rm kd}=1\,{\rm MeV})$, and the blue region is disfavored due to large $Z'$ corrections to the mass splitting.  The brown (purple) region is excluded by BaBar visible (invisible) searches, whereas yellow, pink, gray, green dashed and orange dashed lines are ruled out by NA64, Orsay, E137, SHiP (projected) and Belle-2 (projected), respectively. We have taken $\Delta m=4\,{\rm keV}$ and ${\bar\sigma}_e/m_{\chi_1}=1.2\times 10^{-43}\,{\rm cm}^2/{\rm GeV}$. }
  \label{bounds}
\end{figure}

In Fig.~\ref{bounds}, we depict various constraints on the model in the parameter space for $\varepsilon$ vs $m_{Z'}$ (left in top), $g_{Z'}$ vs $m_{\tilde\pi}$ (right in top) and $m_{\tilde\pi}$ vs $g_{Z'}$ (bottom). First, the XENON1T electron excess \cite{xenon} can be explained along the red line for $\Delta m/\sqrt{3}=2.3\,{\rm keV}$ (then, $\Delta m=4.0\,{\rm keV}$) and ${\bar\sigma}_e/m_{\chi_1}=1.2\times 10^{-43}\,{\rm cm}^2/{\rm GeV}$, whereas the correct relic density is saturated along the red line. The late decoupling condition for $T_{\rm kd}<1\,{\rm keV}$ rules out the gray region, because the relic densities of heavier dark mesons are exponentially suppressed. We also show the stronger bound from the late decoupling condition for $T_{\rm kd}=1\,{\rm MeV}$ in gray line. The blue region is disfavored because of large $Z'$ corrections to the mass splitting between dark mesons. We note that the relic density can be saturated being consistent with XENON1T excess and other experimental bounds, near the resonance with $m_{Z'}=m_{h'}=2m_{\tilde\pi}$ or when the forbidden channels are relevant near $m_{Z'}\gtrsim m_{\tilde\pi}$. These are clearly shown in sharp features of the relic density lines in Fig.~\ref{bounds}.

As shown in Fig.~\ref{bounds}, BaBar visible \cite{babar-vis} and invisible \cite{babar-inv} searches exclude the region in brown and purple regions, and NA64 \cite{NA64}, E137 \cite{e137}, Orsay \cite{orsay} beam dump experiments also constrain the parameter space more strongly and complementarily below about $m_{Z'}=200\,{\rm MeV}$. We note that the bounds from NA64 are stronger than those from E141 \cite{e141} and  E774 \cite{e774}, which are not shown in Fig.~\ref{bounds}. Projected bounds from Belle-2 \cite{belle2} and SHiP \cite{ship} are also shown in dashed orange line and green lines, respectively.

\subsection{Dark meson self-scattering}

All the dark mesons can survive until present for their long lifetime with small mass splittings.
Then, the heavier dark mesons can self-scatter into the lighter states, but the inverse processes are forbidden for dark matter at galaxies at present, because of the small kinetic energy of dark mesons, $E_{\rm DM}=\frac{1}{2}m_{\rm DM} v^2\sim 0.1\,{\rm keV}$, for $v\sim 10^{-3}$ and $m_{\rm DM}\sim 100\,{\rm MeV}$. 
Thus, taking into account the mass hierarchy between dark mesons, $m_{{\widetilde\eta}^0}>m_{{\widetilde K}^0}>m_{{\widetilde\pi}^\pm}>m_{{\widetilde K}^\pm}>m_{{\widetilde\pi}^0}$ as in the previous section, the number of kinematically allowed self-scattering processes at galaxies is reduced:
${\widetilde\pi}^i {\widetilde\pi}^i\rightarrow{\widetilde \pi}^j {\widetilde\pi}^j$ with $i<j$ where $i,j $ run for ${\widetilde\eta}^0, {\widetilde K}^0, {\widetilde\pi}^\pm, {\widetilde K}^\pm, {\widetilde\pi}^0$, and ${\widetilde\pi}^i{\widetilde\pi}^j\rightarrow {\widetilde\pi}^i {\widetilde\pi}^j$ with $i,j$ over all dark mesons, etc.

As a result, the effective self-scattering cross section for dark mesons with split masses at galaxies is given by
\bea
\sigma_{\rm self}&=& \frac{m^2_{\widetilde{\pi}}}{8192\pi}\bigg(\frac{93}{f_\pi^4}+\frac{416 g_{Z'}^2}{f_\pi^2m_{Z'}^2}+\frac{4032 g_{Z'}^4}{m_{Z'}^4}\bigg)
\eea
In our model, due to the strong limit on the dark matter self-annihilation from late decoupling, given in eqs.~(\ref{latedec1}) and (\ref{latedec2}), the expansion parameter for dark chiral perturbation theory should be $m_{\tilde\pi}/f_\pi\lesssim 1$, so the self-interactions for dark mesons give a small contribution to the self-scattering cross section.
For instance, for $m_{\tilde\pi}/f_\pi\sim 0.1, m_{\tilde\pi}\sim 130\,{\rm MeV}$, $g_{Z'}\sim 0.01$ and $m_{Z'}=100\,{\rm MeV}$, which gives rise to a correct relic density being consistent with other constraints,  the self-scattering cross section per unit dark matter is about $\sigma_{\rm self}/m_{\tilde\pi}\sim 4\times 10^{-8}\,{\rm cm^2/g}$, which is too small to give an observable signature at the moment.

\subsection{Lifetime of dark mesons}

For $N_f=3$, we have the mass hierarchy, $m_{{\widetilde\eta}^0}>m_{{\widetilde K}^0}>m_{{\widetilde\pi}^\pm}>m_{{\widetilde K}^\pm}>m_{{\widetilde\pi}^0}$. 
Then, the available decay rates for the heavier dark mesons are 
\bea
\Gamma({\widetilde K}^0\rightarrow {\widetilde K}^+\nu{\bar \nu})&\simeq & \Gamma(\overline{{\widetilde K}^0}\rightarrow {\widetilde K}^-\nu{\bar \nu})
\simeq \frac{ N_\nu e^2 \varepsilon^2 g^2_{Z'}(\Delta m)^5}{480 \pi^3 c_W^4 m^4_Z} \nonumber \\
&=&(9.1\times 10^{25}\,{\rm sec} )^{-1}\bigg(\frac{\varepsilon}{10^{-4}}\bigg)^2\bigg(\frac{g_{Z'}}{0.1}\bigg)^2\bigg(\frac{\Delta m}{4\,{\rm keV}}\bigg)^5, \\
\Gamma({\widetilde \pi}^\pm\rightarrow {\widetilde \pi}^0\nu{\bar \nu}) &\simeq&  \frac{ N_\nu e^2 \varepsilon^2 g^2_{Z'} (\Delta m)^5}{2160\sqrt{3} \pi^3 c_W^4 m^4_Z} \nonumber \\
&\simeq & (7.1\times 10^{26}\,{\rm sec} )^{-1}\bigg(\frac{\varepsilon}{10^{-4}}\bigg)^2\bigg(\frac{g_{Z'}}{0.1}\bigg)^2\bigg(\frac{\Delta m}{4\,{\rm keV}}\bigg)^5.
\eea
Therefore, the heavier mesons are long lived in the parameter space for explaining the Xenon excess and survive until the current Universe. On the other hand, the lighter dark mesons are absolutely stable.

\section{Conclusions}

We presented the possibility that the electron excess reported by XENON1T can be explained by the exothermic scattering between dark mesons with split masses and electron. The flavor-dependent $U(1)'$ interactions for dark quarks are introduced to communicate between dark mesons and the SM through $Z'$ portal, and small mass splittings for dark mesons are generated due to mixing Yukawa couplings for dark Higgs after the $U(1)'$ is broken spontaneously. A small $U(1)'$ gauge coupling and a relatively heavy $Z'$ mass ensure the radiative stability of the mass splitting between dark mesons, in turn, the longevity of dark mesons. We have made the general discussion on split dark mesons and important model-independent constraints in light of the XENON1T excess.

Focusing on the case with three light dark quarks respecting the approximate flavor symmetry, we showed that there are two resolvable monochromatic peaks in the electron recoil spectrum, as a result of the inelastic scattering between dark mesons and electron inside Xenon atom, and our results indicate a better fit to XENON1T data as compared to the case with one monochromatic peak.
We found that there exists a viable parameter space for explaining the XENON1T excess and satisfying various conditions from the radiative stability of the mass splitting for dark mesons, the late chemical decoupling, the correct relic density, as well as various experimental bounds from light $Z'$ searches.
More parameter space is opening up for a low kinetic decoupling temperature as  in the case with extra light fermions in the dark sector or in the presence of a cancellation mechanism of  radiative $Z'$ corrections to split masses for dark mesons. 
We leave a further discussion on those important issues  in a future work.

\section*{Acknowledgments}

We would like to thank Jongkuk Kim, Pyungwon Ko, Jong Chul Park and Seodong Shin for comments and discussion on the related topics. 
This work of SMC was supported in part by the DFG Collaborative Research Centre ``Neutrinos and Dark Matter in Astro- and Particle Physics" (SFB 1258).
The work of HML is supported in part by Basic Science Research Program
through the National Research Foundation of Korea (NRF) funded by the
Ministry of Education, Science and Technology (NRF-2019R1A2C2003738 and NRF-2018R1A4A1025334).
The work of BZ is supported partially by Korea Research Fellowship Program through the National Research Foundation of Korea (NRF) funded by the Ministry of Science and ICT (2019H1D3A1A01070937).

\appendices%

\section{Wess-Zumino-Witten $Z'$ interactions for dark mesons \label{app:Boltz}}

For dark flavors with $N_f=3$, the gauged WZW terms contain the $Z'$ couplings to three dark mesons  are  enumerated as
\bea
{\cal L}_{Z',3\pi} &=&-\frac{i N_c g_{Z'}}{12\sqrt{3} \pi^2 f^3_\pi}\, \epsilon^{\mu\nu\rho\sigma} Z'_\mu \partial_\nu \eta^0\partial_\rho K^+ \partial_\sigma K^--\frac{3 i N_c g_{Z'}}{4\sqrt{3} \pi^2 f^3_\pi}\, \epsilon^{\mu\nu\rho\sigma} Z'_\mu\partial_\nu \eta^0\partial_\rho K^0 \partial_\sigma \overline{K^0}  \nonumber \\
&&+\frac{i N_c g_{Z'}}{6\sqrt{3} \pi^2 f^3_\pi}\, \epsilon^{\mu\nu\rho\sigma} Z'_\mu \partial_\nu \eta^0\partial_\rho \pi^+ \partial_\sigma \pi^- \nonumber \\
&&+\frac{i N_c g_{Z'}}{12 \pi^2 f^3_\pi}\, \epsilon^{\mu\nu\rho\sigma} Z'_\mu\partial_\nu \pi^0\partial_\rho K^+ \partial_\sigma K^-  +\frac{i N_c g_{Z'}}{4 \pi^2 f^3_\pi}\, \epsilon^{\mu\nu\rho\sigma} Z'_\mu\partial_\nu \pi^0\partial_\rho K^0 \partial_\sigma \overline{K^0} \nonumber \\
&& -\frac{i\sqrt{2} N_c g_{Z'}}{12 \pi^2 f^3_\pi}\, \epsilon^{\mu\nu\rho\sigma} Z'_\mu\partial_\nu \pi^+\partial_\rho K^0 \partial_\sigma K^-\nonumber \\
&& -\frac{ i \sqrt{2} N_c g_{Z'}}{12 \pi^2 f^3_\pi}\, \epsilon^{\mu\nu\rho\sigma} Z'_\mu\partial_\nu \pi^-\partial_\rho K^+ \partial_\sigma \overline{K^0}. \label{3pi}
\eea
If  the dark flavor mixings are included,  the above gauged WZW terms become
\bea
{\cal L}_{Z',3{\widetilde\pi}} &=&-\frac{5 i N_c g_{Z'}}{12\sqrt{3} \pi^2 f^3_\pi}\, \epsilon^{\mu\nu\rho\sigma} Z'_\mu \partial_\nu {\widetilde\eta}^0\partial_\rho {\widetilde K}^+ \partial_\sigma {\widetilde K}^--\frac{5 i N_c g_{Z'}}{12\sqrt{3} \pi^2 f^3_\pi}\, \epsilon^{\mu\nu\rho\sigma} Z'_\mu\partial_\nu {\widetilde\eta}^0\partial_\rho {\widetilde K}^0 \partial_\sigma \overline{{\widetilde K}^0}  \nonumber \\
&&-\frac{i N_c g_{Z'}}{6\sqrt{6} \pi^2 f^3_\pi}\, \epsilon^{\mu\nu\rho\sigma} Z'_\mu \partial_\nu {\widetilde\eta}^0\partial_\rho {\widetilde\pi}^0 \partial_\sigma{\widetilde \pi}^- +\frac{i N_c g_{Z'}}{6\sqrt{6} \pi^2 f^3_\pi}\, \epsilon^{\mu\nu\rho\sigma} Z'_\mu \partial_\nu {\widetilde\eta}^0\partial_\rho {\widetilde\pi}^0 \partial_\sigma {\widetilde\pi}^+ \nonumber \\
&&+\frac{i N_c g_{Z'}}{3\sqrt{3} \pi^2 f^3_\pi}\, \epsilon^{\mu\nu\rho\sigma} Z'_\mu \partial_\nu{\widetilde\eta}^0 \partial_\rho {\widetilde K}^0\partial_\sigma {\widetilde K}^- +\frac{i N_c g_{Z'}}{3\sqrt{3} \pi^2 f^3_\pi}\, \epsilon^{\mu\nu\rho\sigma} Z'_\mu \partial_\nu {\widetilde\eta}^0 \partial_\rho {\widetilde K}^+\partial_\sigma \overline{{\widetilde K}^0} \nonumber \\ 
&&-\frac{i N_c g_{Z'}}{12 \pi^2 f^3_\pi}\, \epsilon^{\mu\nu\rho\sigma} Z'_\mu\partial_\nu {\widetilde\pi}^0\partial_\rho {\widetilde K}^+ \partial_\sigma {\widetilde K}^-  +\frac{i N_c g_{Z'}}{12 \pi^2 f^3_\pi}\, \epsilon^{\mu\nu\rho\sigma} Z'_\mu\partial_\nu {\widetilde \pi}^0\partial_\rho {\widetilde K}^0 \partial_\sigma \overline{{\widetilde K}^0}  \\ 
&& +\frac{i\sqrt{2} N_c g_{Z'}}{12 \pi^2 f^3_\pi}\, \epsilon^{\mu\nu\rho\sigma} Z'_\mu\partial_\nu {\widetilde \pi}^+\partial_\rho {\widetilde K}^0 \partial_\sigma \overline{{\widetilde K}^0}  +\frac{i\sqrt{2} N_c g_{Z'}}{12 \pi^2 f^3_\pi}\, \epsilon^{\mu\nu\rho\sigma} Z'_\mu\partial_\nu {\widetilde\pi}^-\partial_\rho {\widetilde K}^0 \partial_\sigma \overline{{\widetilde K}^0} 
\nonumber \\
&& -\frac{i\sqrt{2} N_c g_{Z'}}{12 \pi^2 f^3_\pi}\, \epsilon^{\mu\nu\rho\sigma} Z'_\mu\partial_\nu {\widetilde\pi}^+\partial_\rho {\widetilde K}^0 \partial_\sigma {\widetilde K}^--\frac{ i \sqrt{2} N_c g_{Z'}}{12 \pi^2 f^3_\pi}\, \epsilon^{\mu\nu\rho\sigma} Z'_\mu\partial_\nu {\widetilde \pi}^-\partial_\rho {\widetilde K}^+ \partial_\sigma \overline{{\widetilde K}^0} \nonumber \\
&&+\frac{i\sqrt{2} N_c g_{Z'}}{12 \pi^2 f^3_\pi}\, \epsilon^{\mu\nu\rho\sigma} Z'_\mu\partial_\nu {\widetilde\pi}^+\partial_\rho {\widetilde K}^+ \partial_\sigma {\widetilde K}^-+\frac{ i \sqrt{2} N_c g_{Z'}}{12 \pi^2 f^3_\pi}\, \epsilon^{\mu\nu\rho\sigma} Z'_\mu\partial_\nu {\widetilde \pi}^-\partial_\rho {\widetilde K}^+ \partial_\sigma {\widetilde K}^-. \nonumber 
\eea


\begin{thebibliography}{999}



\bibitem{xenon}
%\cite{Aprile:2020tmw}
%\bibitem{Aprile:2020tmw}
E.~Aprile \textit{et al.} [XENON],
%``Excess electronic recoil events in XENON1T,''
Phys. Rev. D \textbf{102} (2020) no.7, 072004
doi:10.1103/PhysRevD.102.072004
[arXiv:2006.09721 [hep-ex]].
%165 citations counted in INSPIRE as of 06 Dec 2020



\bibitem{exothermic}
%\cite{Batell:2009vb}
%\bibitem{Batell:2009vb}
B.~Batell, M.~Pospelov and A.~Ritz,
%``Direct Detection of Multi-component Secluded WIMPs,''
Phys. Rev. D \textbf{79} (2009), 115019
doi:10.1103/PhysRevD.79.115019
[arXiv:0903.3396 [hep-ph]];
%108 citations counted in INSPIRE as of 11 Oct 2020
%\cite{Graham:2010ca}
%\bibitem{Graham:2010ca}
P.~W.~Graham, R.~Harnik, S.~Rajendran and P.~Saraswat,
%``Exothermic Dark Matter,''
Phys. Rev. D \textbf{82} (2010), 063512
doi:10.1103/PhysRevD.82.063512
[arXiv:1004.0937 [hep-ph]];
%107 citations counted in INSPIRE as of 22 Jun 2020
%\cite{Essig:2010ye}
%\bibitem{Essig:2010ye}
R.~Essig, J.~Kaplan, P.~Schuster and N.~Toro,
%``On the Origin of Light Dark Matter Species,''
[arXiv:1004.0691 [hep-ph]].
%100 citations counted in INSPIRE as of 26 Jun 2020
%\cite{Bernal:2017mqb}
%\bibitem{Bernal:2017mqb}
N.~Bernal, X.~Chu and J.~Pradler,
%``Simply split strongly interacting massive particles,''
Phys. Rev. D \textbf{95} (2017) no.11, 115023
doi:10.1103/PhysRevD.95.115023
[arXiv:1702.04906 [hep-ph]].
%40 citations counted in INSPIRE as of 07 Dec 2020


\bibitem{exodm}
%\cite{Lee:2020wmh}
%\bibitem{Lee:2020wmh}
H.~M.~Lee,
%``Exothermic Dark Matter for XENON1T Excess,''
[arXiv:2006.13183 [hep-ph]].
%39 citations counted in INSPIRE as of 06 Dec 2020


\bibitem{harigaya}
%\cite{Harigaya:2020ckz}
%\bibitem{Harigaya:2020ckz}
K.~Harigaya, Y.~Nakai and M.~Suzuki,
%``Inelastic Dark Matter Electron Scattering and the XENON1T Excess,''
Phys. Lett. B \textbf{809} (2020), 135729
doi:10.1016/j.physletb.2020.135729
[arXiv:2006.11938 [hep-ph]].
%45 citations counted in INSPIRE as of 05 Dec 2020


\bibitem{essig}
%\cite{Bloch:2020uzh}
%\bibitem{Bloch:2020uzh}
I.~M.~Bloch, A.~Caputo, R.~Essig, D.~Redigolo, M.~Sholapurkar and T.~Volansky,
%``Exploring New Physics with O(keV) Electron Recoils in Direct Detection Experiments,''
[arXiv:2006.14521 [hep-ph]].
%56 citations counted in INSPIRE as of 28 Nov 2020


\bibitem{exodmothers}
%\cite{Bramante:2020zos}
%\bibitem{Bramante:2020zos}
J.~Bramante and N.~Song,
%``Electric But Not Eclectic: Thermal Relic Dark Matter for the XENON1T Excess,''
Phys. Rev. Lett. \textbf{125} (2020) no.16, 161805
doi:10.1103/PhysRevLett.125.161805
[arXiv:2006.14089 [hep-ph]];
%40 citations counted in INSPIRE as of 05 Dec 2020
%\cite{Baek:2020owl}
%\bibitem{Baek:2020owl}
S.~Baek, J.~Kim and P.~Ko,
%``XENON1T excess in local $Z_2$ DM models with light dark sector,''
Phys. Lett. B \textbf{810} (2020), 135848
doi:10.1016/j.physletb.2020.135848
[arXiv:2006.16876 [hep-ph]];
%\cite{Borah:2020jzi}
%\bibitem{Borah:2020jzi}
D.~Borah, S.~Mahapatra, D.~Nanda and N.~Sahu,
%``Inelastic fermion dark matter origin of XENON1T excess with muon $(g ???2)$ and light neutrino mass,''
Phys. Lett. B \textbf{811} (2020), 135933
doi:10.1016/j.physletb.2020.135933
[arXiv:2007.10754 [hep-ph]];
%10 citations counted in INSPIRE as of 13 Dec 2020
%25 citations counted in INSPIRE as of 05 Dec 2020
%\cite{Aboubrahim:2020iwb}
%\bibitem{Aboubrahim:2020iwb}
A.~Aboubrahim, M.~Klasen and P.~Nath,
%``Xenon-1T excess as a possible signal of a sub-GeV hidden sector dark matter,''
[arXiv:2011.08053 [hep-ph]].
%1 citations counted in INSPIRE as of 13 Dec 2020


\bibitem{review}
%\cite{Lee:2020eap}
%\bibitem{Lee:2020eap}
H.~M.~Lee,
%``Dark mesons as self-interacting dark matter,''
[arXiv:2008.13090 [hep-ph]].
%0 citations counted in INSPIRE as of 06 Dec 2020


\bibitem{simpmeson2}
%\cite{Lee:2015gsa}
%\bibitem{Lee:2015gsa}
H.~M.~Lee and M.~S.~Seo,
%``Communication with SIMP dark mesons via Z? -portal,''
Phys. Lett. B \textbf{748} (2015), 316-322
doi:10.1016/j.physletb.2015.07.013
[arXiv:1504.00745 [hep-ph]].
%56 citations counted in INSPIRE as of 21 Jul 2020



\bibitem{wz}
%\cite{Wess:1971yu}
%\bibitem{Wess:1971yu}
J.~Wess and B.~Zumino,
%``Consequences of anomalous Ward identities,''
Phys. Lett. B \textbf{37} (1971), 95-97
doi:10.1016/0370-2693(71)90582-X
%2713 citations counted in INSPIRE as of 27 Jul 2020


\bibitem{witten}
%\cite{Witten:1983tw}
%\bibitem{Witten:1983tw}
E.~Witten,
%``Global Aspects of Current Algebra,''
Nucl. Phys. B \textbf{223} (1983), 422-432
doi:10.1016/0550-3213(83)90063-9
%2709 citations counted in INSPIRE as of 27 Jul 2020


\bibitem{simpmeson1}
%\cite{Hochberg:2014kqa}
%\bibitem{Hochberg:2014kqa}
Y.~Hochberg, E.~Kuflik, H.~Murayama, T.~Volansky and J.~G.~Wacker,
%``Model for Thermal Relic Dark Matter of Strongly Interacting Massive Particles,''
Phys. Rev. Lett. \textbf{115} (2015) no.2, 021301
doi:10.1103/PhysRevLett.115.021301
[arXiv:1411.3727 [hep-ph]].
%184 citations counted in INSPIRE as of 21 Jul 2020





\bibitem{simpmeson3}
%\cite{Choi:2018iit}
%\bibitem{Choi:2018iit}
S.~M.~Choi, H.~M.~Lee, P.~Ko and A.~Natale,
%``Resolving phenomenological problems with strongly-interacting-massive-particle models with dark vector resonances,''
Phys. Rev. D \textbf{98} (2018) no.1, 015034
doi:10.1103/PhysRevD.98.015034
[arXiv:1801.07726 [hep-ph]].
%25 citations counted in INSPIRE as of 21 Jul 2020


\bibitem{gori}
%\cite{Berlin:2018tvf}
%\bibitem{Berlin:2018tvf}
A.~Berlin, N.~Blinov, S.~Gori, P.~Schuster and N.~Toro,
%``Cosmology and Accelerator Tests of Strongly Interacting Dark Matter,''
Phys. Rev. D \textbf{97} (2018) no.5, 055033
doi:10.1103/PhysRevD.97.055033
[arXiv:1801.05805 [hep-ph]].
%48 citations counted in INSPIRE as of 06 Dec 2020



\bibitem{split}
%\cite{Katz:2020ywn}
%\bibitem{Katz:2020ywn}
A.~Katz, E.~Salvioni and B.~Shakya,
%``Split SIMPs with Decays,''
JHEP \textbf{10} (2020), 049
doi:10.1007/JHEP10(2020)049
[arXiv:2006.15148 [hep-ph]];
%2 citations counted in INSPIRE as of 05 Dec 2020
%\cite{Balkin:2018tma}
%\bibitem{Balkin:2018tma}
R.~Balkin, M.~Ruhdorfer, E.~Salvioni and A.~Weiler,
%``Dark matter shifts away from direct detection,''
JCAP \textbf{11} (2018), 050
doi:10.1088/1475-7516/2018/11/050
[arXiv:1809.09106 [hep-ph]].
%25 citations counted in INSPIRE as of 05 Dec 2020




\bibitem{z3dm}
%\cite{Choi:2015bya}
%\bibitem{Choi:2015bya}
S.~M.~Choi and H.~M.~Lee,
%``SIMP dark matter with gauged Z$_{3}$ symmetry,''
JHEP \textbf{09} (2015), 063
doi:10.1007/JHEP09(2015)063
[arXiv:1505.00960 [hep-ph]].
%58 citations counted in INSPIRE as of 24 Jul 2020




\bibitem{dashen}
%\cite{Dashen:1969eg}
%\bibitem{Dashen:1969eg}
R.~F.~Dashen,
%``Chiral SU(3) x SU(3) as a symmetry of the strong interactions,''
Phys. Rev. \textbf{183} (1969), 1245-1260
doi:10.1103/PhysRev.183.1245
%719 citations counted in INSPIRE as of 05 Dec 2020





\bibitem{formfactor}
%\cite{Essig:2019xkx}
%\bibitem{Essig:2019xkx}
R.~Essig, J.~Pradler, M.~Sholapurkar and T.~T.~Yu,
%``Relation between the Migdal Effect and Dark Matter-Electron Scattering in Isolated Atoms and Semiconductors,''
Phys. Rev. Lett. \textbf{124} (2020) no.2, 021801
doi:10.1103/PhysRevLett.124.021801
[arXiv:1908.10881 [hep-ph]].
%24 citations counted in INSPIRE as of 04 Dec 2020




\bibitem{vsimp}
%\cite{Choi:2017zww}
%\bibitem{Choi:2017zww}
S.~M.~Choi, Y.~Hochberg, E.~Kuflik, H.~M.~Lee, Y.~Mambrini, H.~Murayama and M.~Pierre,
%``Vector SIMP dark matter,''
JHEP \textbf{10} (2017), 162
doi:10.1007/JHEP10(2017)162
[arXiv:1707.01434 [hep-ph]];
%45 citations counted in INSPIRE as of 28 Nov 2020
%\cite{Choi:2019zeb}
%\bibitem{Choi:2019zeb}
S.~M.~Choi, H.~M.~Lee, Y.~Mambrini and M.~Pierre,
%``Vector SIMP dark matter with approximate custodial symmetry,''
JHEP \textbf{07} (2019), 049
doi:10.1007/JHEP07(2019)049
[arXiv:1904.04109 [hep-ph]].
%12 citations counted in INSPIRE as of 28 Nov 2020



\bibitem{latedec}
%\cite{Aarssen:2012fx}
%\bibitem{Aarssen:2012fx}
L.~G.~van den Aarssen, T.~Bringmann and C.~Pfrommer,
%``Is dark matter with long-range interactions a solution to all small-scale problems of \textbackslash{}Lambda CDM cosmology?,''
Phys. Rev. Lett. \textbf{109} (2012), 231301
doi:10.1103/PhysRevLett.109.231301
[arXiv:1205.5809 [astro-ph.CO]];
%191 citations counted in INSPIRE as of 30 Nov 2020
%\cite{Bringmann:2013vra}
%\bibitem{Bringmann:2013vra}
T.~Bringmann, J.~Hasenkamp and J.~Kersten,
%``Tight bonds between sterile neutrinos and dark matter,''
JCAP \textbf{07} (2014), 042
doi:10.1088/1475-7516/2014/07/042
[arXiv:1312.4947 [hep-ph]].
%94 citations counted in INSPIRE as of 30 Nov 2020



\bibitem{babar-vis}
%\cite{Lees:2014xha}
%\bibitem{Lees:2014xha}
J.~Lees \textit{et al.} [BaBar],
%``Search for a Dark Photon in $e^+e^-$ Collisions at BaBar,''
Phys. Rev. Lett. \textbf{113} (2014) no.20, 201801
doi:10.1103/PhysRevLett.113.201801
[arXiv:1406.2980 [hep-ex]].
%335 citations counted in INSPIRE as of 23 Jun 2020


\bibitem{babar-inv}
%\cite{Lees:2017lec}
%\bibitem{Lees:2017lec}
J.~Lees \textit{et al.} [BaBar],
%``Search for Invisible Decays of a Dark Photon Produced in ${e}^{+}{e}^{-}$ Collisions at BaBar,''
Phys. Rev. Lett. \textbf{119} (2017) no.13, 131804
doi:10.1103/PhysRevLett.119.131804
[arXiv:1702.03327 [hep-ex]].
%178 citations counted in INSPIRE as of 23 Jun 2020


\bibitem{NA64}
%\cite{NA64:2019imj}
%\bibitem{NA64:2019imj}
D.~Banerjee, V.~E.~Burtsev, A.~G.~Chumakov, D.~Cooke, P.~Crivelli, E.~Depero, A.~V.~Dermenev, S.~V.~Donskov, R.~R.~Dusaev and T.~Enik, \textit{et al.}
%``Dark matter search in missing energy events with NA64,''
Phys. Rev. Lett. \textbf{123} (2019) no.12, 121801
doi:10.1103/PhysRevLett.123.121801
[arXiv:1906.00176 [hep-ex]].
%58 citations counted in INSPIRE as of 11 Oct 2020


\bibitem{e137}
%\cite{Bjorken:1988as}
%\bibitem{Bjorken:1988as}
J.~D.~Bjorken, S.~Ecklund, W.~R.~Nelson, A.~Abashian, C.~Church, B.~Lu, L.~W.~Mo, T.~A.~Nunamaker and P.~Rassmann,
%``Search for Neutral Metastable Penetrating Particles Produced in the SLAC Beam Dump,''
Phys. Rev. D \textbf{38} (1988), 3375
doi:10.1103/PhysRevD.38.3375
%314 citations counted in INSPIRE as of 03 Mar 2021


\bibitem{orsay}
%\cite{Davier:1989wz}
%\bibitem{Davier:1989wz}
M.~Davier and H.~Nguyen Ngoc,
%``An Unambiguous Search for a Light Higgs Boson,''
Phys. Lett. B \textbf{229} (1989), 150-155
doi:10.1016/0370-2693(89)90174-3
%162 citations counted in INSPIRE as of 03 Mar 2021


\bibitem{e141}
%\cite{Riordan:1987aw}
%\bibitem{Riordan:1987aw}
E.~M.~Riordan, M.~W.~Krasny, K.~Lang, P.~De Barbaro, A.~Bodek, S.~Dasu, N.~Varelas, X.~Wang, R.~G.~Arnold and D.~Benton, \textit{et al.}
%``A Search for Short Lived Axions in an Electron Beam Dump Experiment,''
Phys. Rev. Lett. \textbf{59} (1987), 755
doi:10.1103/PhysRevLett.59.755
%291 citations counted in INSPIRE as of 03 Mar 2021


\bibitem{e774}
%\cite{Bross:1989mp}
%\bibitem{Bross:1989mp}
A.~Bross, M.~Crisler, S.~H.~Pordes, J.~Volk, S.~Errede and J.~Wrbanek,
%``A Search for Shortlived Particles Produced in an Electron Beam Dump,''
Phys. Rev. Lett. \textbf{67} (1991), 2942-2945
doi:10.1103/PhysRevLett.67.2942
%204 citations counted in INSPIRE as of 03 Mar 2021


\bibitem{belle2}
%\cite{Essig:2013vha}
%\bibitem{Essig:2013vha}
R.~Essig, J.~Mardon, M.~Papucci, T.~Volansky and Y.~M.~Zhong,
%``Constraining Light Dark Matter with Low-Energy $e^+e^-$ Colliders,''
JHEP \textbf{11} (2013), 167
doi:10.1007/JHEP11(2013)167
[arXiv:1309.5084 [hep-ph]].
%163 citations counted in INSPIRE as of 30 Nov 2020


\bibitem{ship}
%\cite{SHiP:2020noy}
%\bibitem{SHiP:2020noy}
C.~Ahdida \textit{et al.} [SHiP],
%``Sensitivity of the SHiP experiment to light dark matter,''
[arXiv:2010.11057 [hep-ex]].
%1 citations counted in INSPIRE as of 30 Nov 2020




\end{thebibliography}
\end{document}